\documentclass[prb,twocolumn,showpacs,preprintnumbers,amsmath,amssymb,superscriptaddress]{revtex4-2}
\usepackage[T1]{fontenc} 
\usepackage{graphicx}% Include figure files
\usepackage{dcolumn}% Align table columns on decimal point
\usepackage{bm}% bold math
\usepackage{color}
\usepackage{multirow}
\usepackage{braket}
\usepackage{here}
\usepackage{ulem}
\usepackage{xcolor}
\DeclareRobustCommand{\erase}{\bgroup\markoverwith{\textcolor{red}{\rule[.5ex]{2pt}{0.4pt}}}\ULon}

\begin{document}

\title{Experimental nuclear quadrupole resonance and computational study of the structurally refined topological semimetal TaSb$_2$.
 }

\author{T. Fujii}
\affiliation{Max Planck Institute for Chemical Physics of Solids, Nothnitzer Stra{\ss}e 40, 01187 Dresden, Germany}
\email{Takuto.Fujii@cpfs.mpg.de}
\author{O. Janson}
\affiliation{Leibniz Institute for Solid State and Materials Research IFW Dresden, 01069 Dresden, Germany}
\author{H. Yasuoka}
\affiliation{Max Planck Institute for Chemical Physics of Solids, Nothnitzer Stra{\ss}e 40, 01187 Dresden, Germany}
\author{H. Rosner}
\affiliation{Max Planck Institute for Chemical Physics of Solids, Nothnitzer Stra{\ss}e 40, 01187 Dresden, Germany}
\author{Yu. Prots}
\affiliation{Max Planck Institute for Chemical Physics of Solids, Nothnitzer Stra{\ss}e 40, 01187 Dresden, Germany}
\author{U. Burkhardt}
\affiliation{Max Planck Institute for Chemical Physics of Solids, Nothnitzer Stra{\ss}e 40, 01187 Dresden, Germany}
\author{M. Schmidt}
\affiliation{Max Planck Institute for Chemical Physics of Solids, Nothnitzer Stra{\ss}e 40, 01187 Dresden, Germany}
\author{M. Baenitz}
\affiliation{Max Planck Institute for Chemical Physics of Solids, Nothnitzer Stra{\ss}e 40, 01187 Dresden, Germany}

\date{\today}

\begin{abstract}
The local electric field gradients and magnetic dynamics of TaSb$_2$ have been studied using
$^{121}$Sb, $^{123}$Sb, and $^{181}$Ta nuclear quadrupole resonance (NQR) with
density functional theory (DFT) calculations using XRD-determined crystal structures. By measuring all structurally expected thirteen NQR
lines, the nuclear quadrupole coupling
constant ($\nu_Q$) and asymmetric parameter ($\eta$) for Ta, Sb(1), and
Sb(2) sites were obtained. These values are all in good agreement with the presented DFT calculations. Principal axes of the electric field gradients
was determined for a single-crystal sample by measuring the
angular dependencies of NMR frequency under a weak magnetic field.
The unusual temperature dependence of $\eta$(T) of Sb(2) hints at the suppressed
thermal expansion along the $a$-axis. Spin-lattice relaxation rate ($1/T_1T$)
measurements reveal an activated-type behavior and an upturn below 30 K. Neither the low temperature upturn nor the high temperature activation type behaviors are reproduced by the calculated $1/T_1T$ based on the calculated density of states (DOS). On the other hand, the agreement between the calculated DOS and specific heat measurements indicates that the band renormalization is small. This fact indicates that TaSb$_2$ deviates from the simple semimetal scenario, and magnetic excitations are not captured by Fermi liquid theory.

\end{abstract}

\maketitle

\section{Introduction}
Topological insulators and semimetals are receiving tremendous attention in the
solid state research community
\cite{Neto,Kane,Bernevig,Hasan2010,Wehling2014,Armitage,Yan2017}. A spontaneous symmetry breaking via spin-orbit interactions generates a new unconventional phenomenon beyond the Fermi Landau theory. This required a completely new concept of theoretical
description and the introduction of new fermions, the so-called Dirac or Weyl
fermions. Dirac and Weyl semimetals have unique electronic structures with linear dispersion in momentum space. The direct experimental proof of this fascinating new physics was
achieved by angle resolved photoemission spectroscopy \cite{Xu2015, lv2015}.
Indirectly one sees signatures of these topological states formed out of Weyl
or Dirac fermions in measurement methods such as resistance (magnetotransport,
Shubnikov-de Haas oscillations \cite{Shekhar,Liang,Narayanan, Huang,Hu}),
magnetization (de Haas-van Alphen oscillations \cite{Klotz2016,Arnold,Besara})
or also in the optical conductivity \cite{Kimura2017,xu2016,Kimura}. The first Weyl semimetals
studied in detail are found among the monopnictides TMPn(TM=Ta,Nb,Pn=P,As)
\cite{Weng2015}. These systems clearly show the presence of the Weyl nodes near
the Fermi level \cite{Huang2015}. After the discovery of superconductivity in Dirac/Weyl monopnictides (TlSb, NbP) \cite{Zhou2021,
Baenitz2019}, Weyl dichalcogenides (MoTe$_2$) \cite{Qi2016} and Dirac
dipnictides (CaSb$_2$) \cite{Ikeda2020,Takahashi2021}, we extend our studies
towards the dipnictides TMPn$_2$(TM=Nb, Ta, Pn= P, As, Sb). In contrast to the
monopnictides, TMPn$_2$ exhibits a centrosymmetric crystal structure. NbAs$_2$ and
TaSb$_2$ show a negative magnetoresistance \cite{shen,li2}, similar to that
observed in TaAs and NbP \cite{Huang, Niemann}. For the latter compound, this has been claimed to originate from the chiral anomalies in the Weyl fermion. 
Theoretically, the band structure calculation shows no Weyl fermions in TaSb$_2$ and there are electron and hole bands which are not crossing each other \cite{CXu, Gresch}.
% XPn$_2$ is theoretically proposed to present a nodal line around the Fermi level by the band structure calculations without spin-orbit coupling (SOC) \cite{CXu}. Once SOC is taken into consideration, the nodal lines gap out, leaving only a pair of discontinuous electron/hole bands across the Fermi surface \cite{CXu, Gresch}.
Gresch \textit{et al}., theoretically proposed
that the effect of the magnetic field induces the Weyl fermions in TMPn$_2$
\cite{Gresch}, that can explain the appearance of negative magnetoresistance.
However, the question arises what characteristics of the band structure are
formed under a zero magnetic field. A key to understand the physics of these
compounds is to examine whether unique topological characters are present under
a zero magnetic field.

\indent Using $^{181}$Ta nuclear quadrupole resonance (NQR) on TaP, we have succeeded for the first time in detecting the excitations of the Weyl fermions near the Fermi level \cite{Yasuoka}. In general, the nuclear magnetic resonance (NMR) method has been successfully applied to Dirac/Weyl fermion material under a magnetic fields \cite{Tian1, Tian2, Yokoo, Antonenko, Wang2020}, and these experimental results are  consistent with theory \cite{Maebashi, okvatovity}. Ta based topological semimetals are particularly suitable as Ta has one of the largest quadrupole moments which facilitates the NQR measurements. Moreover, the NQR method allows us to explore electronic states under a zero magnetic field. In particular, the spin-lattice relaxation rate ($1/T_1$), reflecting the density of states (DOS) in non-magnetic metals and semimetals, has been successfully applied to several Weyl semimetals \cite{Yasuoka, Wang2020}.

\indent The aim of this paper is to explore the local electric field gradient and the magnetic excitations in TaSb$_2$ using mainly the NQR method. Special attention was given to the quality of both the polycrystalline and the single crystal samples used. Especially in semimetals, it  has been known that stoichiometry plays a crucial role. For this reason, detailed investigations by energy dispersive x-ray spectroscopy (EDXS) and 
structural investigation by x-ray scattering were carried out. We conducted a comprehensive NQR study on the two Sb sites and the Ta site. The temperature dependence of the electric field gradient (EFG) and $1/T_1$ at the two Sb sites and the Ta site have been studied in great detail. Here, we have also conducted single crystal NMR measurements under a weak field (WFNMR) to check the site assignment of the observed NQR lines and to obtain the EFG tensor experimentally as well. The experimental NQR parameters thus obtained in our detailed study are in rather good agreement with the results of the theoretical calculation based on density functional theory (DFT).

%---------------Introduction----------------------------------------------------%

%---------------Sample detail----------------------------------------------------%
\section{Crystal structure and sample preparation}
%TaSb$_2$ crystallizes in the centrosymmetric space group $C12/m$ and contains two inequivalent Sb(1) and Sb(2) sites and one Ta site, as shown in Fig.~\ref{fig:str}. The present NQR results clearly distinguish these three sites.
%Single crystal samples used in the present NQR/NMR study were grown by chemical vapor transport reaction using iodine as a transport agent. TaSb$_2$ has first been synthesized by a direct reaction of the elements tantalum (powder 99,98 \% Alfa Aesar) and antimony (powder 99,999 \% Alfa Aesar) at 800 $^{\circ}$C in evacuated fused silica tubes during 10 days. Starting from this microcrystalline powder, TaSb$_2$ crystallized by a chemical transport reaction in a temperature gradient from 1000 $^{\circ}$C (source) to 950 $^{\circ}$C (sink), and a transport additive concentration of 10 mg/cm$^3$ iodine (99,998\% Alfa Aesar). The crystallization experiment was carried out in a horizontally arranged two-zone furnace which was tilted by approximately 10$^{\circ}$. The typical size of crystal obtained are $3\times3\times1$ mm. The powder sample was obtained by crashing single crystal sample. The sample was investigated by x-ray diffraction. The local stoichiometry was investigated by energy dispersive x-ray (EDX) analysis (see Appendix \ref{ApB}).

\begin{figure}[tb]
\centering
\includegraphics[width=1\linewidth]{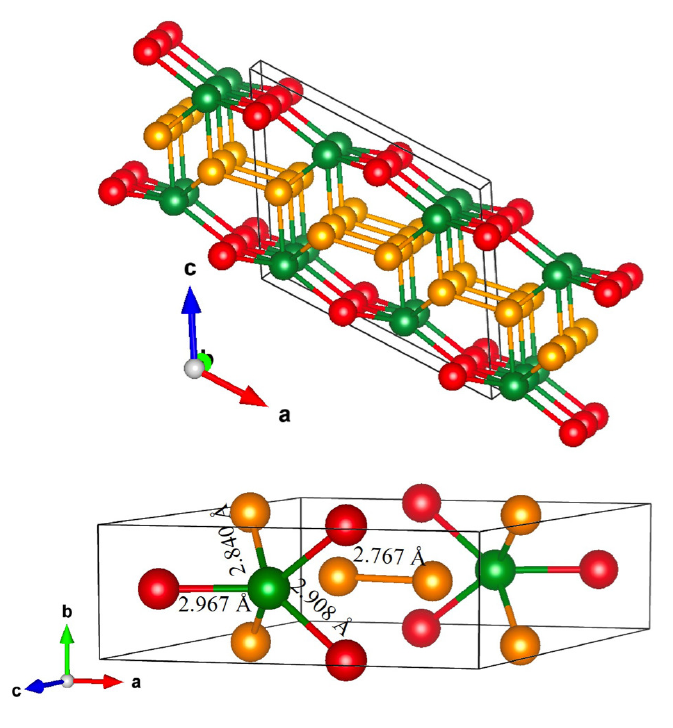}
% Here is how to import EPS art
\caption{\label{fig:str}
Monoclinic crystal structure of TaSb$_2$. There are two inequivalent Sb(1) (red) and Sb(2) (orange) which are bonded to the Ta (green). The cell parameters are $a$=10.233 \AA, $b$=3.648 \AA, $c$=8.303 \AA, $\alpha$=$\gamma$=90.00 $^{\circ}$ and $\beta$=120.40 $^{\circ}$ taken from Appendix \ref{ApA}.%the Inorganic Crystal Structure Database
%(``icsd-651600''~\cite{Hulliger1964}) .  %The atomic positions and Wycoff numbers are listed in table for Ta, Sb(1) and Sb(2) sites
}
\end{figure}

TaSb$_2$ crystallizes in the centrosymmetric space group $C2/m$ (NbSb$_2$-type) and contains two inequivalent Sb(1) and Sb(2) sites, and one Ta site. As shown in Fig. \ref{fig:str}, the skeleton of the structure can be viewed as separate units composed by two Ta and two Sb(1) atoms linked by $P_{\sigma}$ bonded Sb(2) and Sb(2) dumbbells along the $a$-axis (see Fig. \ref{fig:str}). These dumbbells represent a covalent bond between two Sb(2) site. The dumbbell can accommodate four electrons, resulting in an average valence of -2 for the Sb(2) site, while the Sb(1) site has a valence of -3. Thus the Zintl rule is satisfied. 
%Sb(1) site is isolated from each Sb atoms, while Sb(2) site forms dimer of Sb(2)-Sb(2) with shorter bonding than the others along $a$-axis.
%the skeleton of the structure can be viewed as that plackets composed by two Ta and two Sb(2) atoms liked by $P_{\sigma}$ bonded Sb(2) and Sb(2) pairs along the $a$-axis. 
Therefore, it is easily inferred that the electronic and photonic dispersions are highly anisotropic with the $a$-axis being a special axis.

\indent Single crystals were grown by a chemical vapor transport reaction using iodine as a transport agent. TaSb$_2$ has first been synthesized by a direct reaction of the tantalum (powder 99,98 \% Alfa Aesar) and antimony (powder 99,999 \% Alfa Aesar) at 800 $^{\circ}$C in a evacuated silica tube during 10 days. Starting from this microcrystalline powder, TaSb$_2$ crystallized by the chemical transport reaction in a temperature gradient from 1000 $^{\circ}$C (source) to 950 $^{\circ}$C (sink), and a transport additive concentration of 10 mg/cm$^3$ iodine (99,998\% Alfa Aesar). The crystallization experiment was carried out in a horizontally arranged two-zone furnace which was tilted by approximately 10 $^{\circ}$. The typical size of crystal obtained is $3\times3\times1$ mm. The quality of the sample was checked by x-ray diffraction measurement. The local stoichiometry was investigated by energy dispersive x-ray (EDX) analysis (see Appendix \ref{ApA}). 
Both data revealed that the synthesized sample is phase pure and of high homogeneity.
The TaSb$_2$ phase was identified half a century ago \cite{Hulinger} and its structure assigned to the NbSb$_2$ type \cite{Saini}. The roughly atomic coordinates were established from the evaluation of x-ray powder Guinier photographs \cite{Furuseth}. Although a detailed description of the crystal structure of TaSb$_2$ was therefore not yet available, our investigation using x-ray single crystal diffraction have refined detailed crystal structure (see Appendix \ref{ApA}). Hereafter, we will use the refined crystal parameters to analyze NMR data and band structure calculations.

%---------------Sample detail----------------------------------------------------%

%--------------------------Fig.1--------------------------------------------------
%--------------------------Fig.1--------------------------------------------------

%---------------NQR method----------------------------------------------------%

 %--------------------------------------Table.1-----------------------------%
\begin{table}[t]
\caption{\label{tab:nmrnqr}NMR/NQR related nuclear parameters of $^{121}$Sb, $^{123}$Sb, and $^{181}$Ta}
\begin{ruledtabular}
\begin{tabular}{ccddd}
Isotope    & Spin & \multicolumn{1}{c}{$\gamma/2\pi$ (MHz/T)} & \multicolumn{1}{c}{Q (barns)} & \multicolumn{1}{c}{Natural abundance (\%)} \\ \hline
$^{121}$Sb & 5/2  & 10.1890               & -0.54     & 57.2           \\
$^{123}$Sb & 7/2  & 5.5176                & -0.69     & 42.8           \\
$^{181}$Ta & 7/2  & 5.0960                & 3.17      & 100            \\
\end{tabular}
\end{ruledtabular}
\end{table}
%--------------------------------------Table.1-----------------------------%
\section{NQR and NMR techniques}
%TaSb$_2$ contains three NQR/NMR-active nuclei ($^{121}$Sb, $^{123}$Sb and $^{181}$Ta) and two inequivalent Sb sites. The nuclear parameters of $^{121}$Sb, $^{123}$Sb, and $^{181}$Ta are listed in Table~\ref{tab:nmrnqr}. The NQR experiments were carried out on powder (crashed single crystals),  and its spectra were taken by the frequency sweep method. The signal intensities were taken from the real-part echo integral. 
Our target nuclei in TaSb$_2$ are $^{121/123}$Sb and $^{181}$Ta and those nuclear parameters are tabulated in the Table \ref{tab:nmrnqr}. The most prominent feature among those parameters is that $^{181}$Ta has an extraordinary large nuclear quadrupole moment. This immediately suggests that $^{181}$Ta-NQR occurs at a relatively high frequency. The NQR experiments were carried out on powder (crashed single crystals) sample using a conventional pulsed NMR/NQR spectrometer (Tecmag Apollo). The NQR spectra were taken by the frequency-sweep method where the real part of the spin-echo signal was accumulated as a function of frequency.
In order to assign the observed NQR lines to nuclear isotopes and different sites, and to determine the principal axes of the EFG tensor, the weak field NMR (WFNMR) measurements were also carried out using an oriented single crystal, where the resonances were obtained under a weak field (less than 0.05 T) applied to the NQR lines. When a weak field applies along the main principal axis of EFG, the NQR lines splits into two lines. We have also measured the angular dependence of the WFNMR lines from where the principal axes of EFG for two Sb sites and one Ta site were determined. The temperature dependence of nuclear spin-lattice relaxation rate ($1/T_1$) has been measured for the NQR lines by the inversion recovery method. The $T_1$ values are obtained by fitting the recovery of nuclear magnetization, $M(t)$, measured by the spin-echo amplitude after an application of inversion pulse, to the proper recovery function (refer Eqs. (\ref{recovery1}) and (\ref{recovery2}) in Appendix \ref{ApB}).
%---------------NQR method----------------------------------------------------%

\section{DFT band-structure calculations}
First-principles calculations were performed using the full-potential code FPLO
version 21, employing a local-orbital basis set~\cite{FPLO}. To account for the
sizable spin-orbit coupling in TaSb$_2$, we use the full-relativistic
treatment. The local density approximation (LDA) parametrized by Perdew and
Wang~\cite{Pw92} was employed for the exchange-correlation potential. Unless
otherwise mentioned, for the structural input we used the data obtained by our x-ray single crystal diffraction (see Appendix \ref{ApA}).%the data from the
%Inorganic Crystal Structure Database (ICSD code 651600) data from
%Ref.~\cite{Hulliger1964}.

\begin{figure}[tb]
\includegraphics[width=1\linewidth]{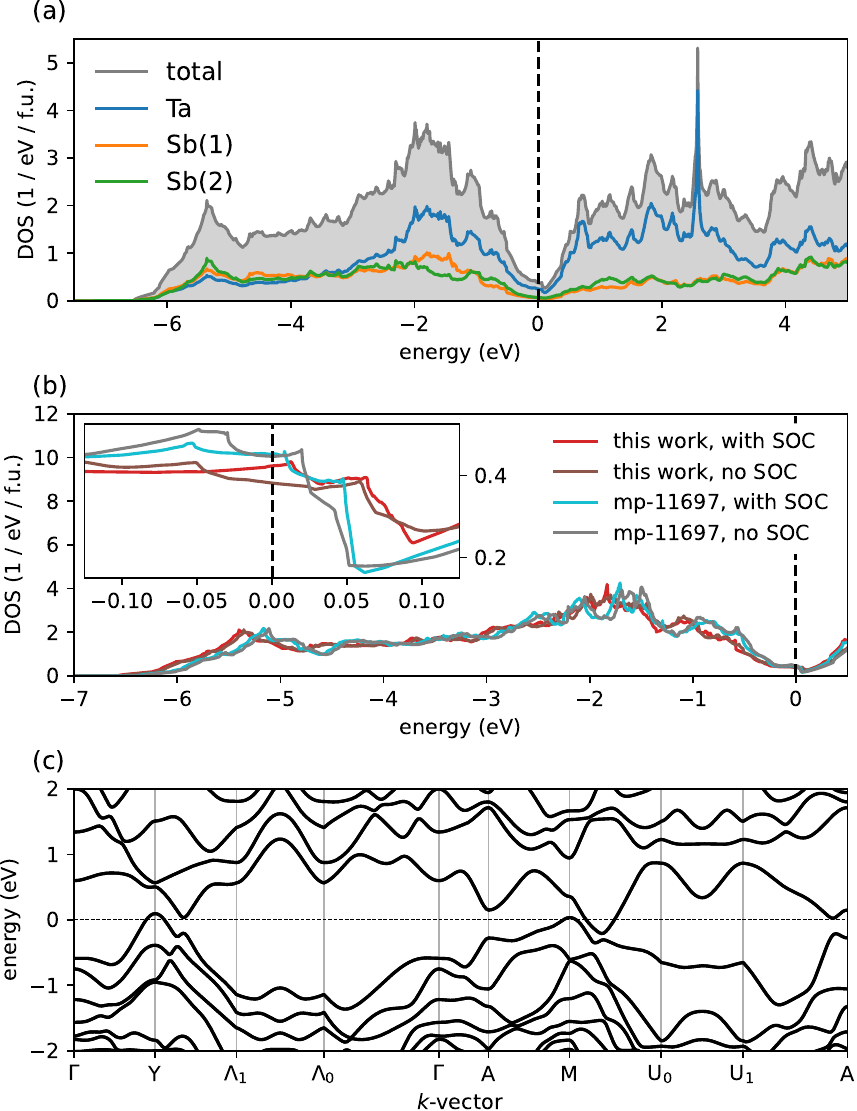}
\caption{\label{fig:dos}
Electronic structure of TaSb$_2$ in nonmagnetic LDA calculations.
(a) Total and site-resolved density of states (DOS) in a full-relativistic
non-mangetic calculation. (b) Comparison of DOS calculated in the 
full relativistic (``with SOC'') and scalar-relativistic (``no SOC'') mode for
two structures: the experimental structure obtained from our x-ray single
crystal diffraction (see Appendix \ref{ApA}) and the optimized structure from the
Materials project (mp-11697~\cite{mp11697}). The inset shows a blow-up of the
same data.  (c) Band structure for the experimental crystal structure. In all
plots, the Fermi level (dashed line) is at zero energy. } \end{figure}

For microscopic insights into the electronic structure of TaSb$_2$, we resort
to the nonmagnetic DOS depicted in Fig.~\ref{fig:dos} (a). 
The valence band
splits into a high-energy manifold dominated by Sb $5p$ states (between $-12.5$
and $-7.5$\,eV) and the low-energy manifold (starting around $-6.5$\,eV and
ranging up to the Fermi energy) to which also Ta $5d$ contributes prominently.
The valence and the polarization bands are bridged by a steep valley with an
intricate structure. Since this exact shape of this valley plays a crucial role
in the transport properties of TaSb$_2$, we resolve it in the inset of
Fig.~\ref{fig:dos} (b). The most prominent feature in this energy range is
the dip centered around 0.1\,eV above the Fermi energy. The precise structure
of the DOS is sensitive to the lattice constants and atomic positions: If we
use the  optimized TaSb$_2$ structural from the Materials project
(mp-11697~\cite{mp11697}), this dip becomes sharper and moves closer to the
Fermi level. Moreover, the difference between scalar-relativistic and full
relativistic calculations reveals a profound impact of the spin-orbit coupling (SOC) on the
electronic structure in this energy range.  Interestingly, if we consider the
full valence band, this difference becomes barely visible  (Fig.~\ref{fig:dos},
(b), main panel).
%Previous studies suggested that TaSb$_2$ is a Weyl semimetal. To study its topological properties, we
%perform full-relativistic GGA calculations and describe the bands near the Fermi level using Wannier projections to Ta 5$d$ and Sb 5$p$ states. 
%In the plotted segments of the Brillouin zone, “HOMO” and “LUMO” bands are separated by a finite gap. To check that this gap exists everywhere in the Brillouin zone, we performed an automatic search for Weyl points using the algorithm implemented in the  find Weyl Points function of “pyfplo.slabify” module. It should be emphasized that no Weyl points were found.

 It has been suggested that a magnetic field may stabilize a Weyl point in TaSb$_2$ \cite{Gresch}. While we can not exclude such scenario, the band structure plot reveals that highest-lying occupied band and the lowest-lying unoccupied band are separated by a finite gap (Fig.~\ref{fig:dos} (c)). Since the magnetic energy scale is roughly set by Bohr magneton which amounts to approximately 5.8$\times$10$^{-5}$ eV/T, closing a gap of 0.1 eV would require an unrealistically large magnetic field.

\section{Results and discussion}

\subsection{NQR spectra and site assignment}

%--------------------------Fig.2--------------------------------------------------
\begin{figure}[t]
\centering
\includegraphics[width=1\linewidth]{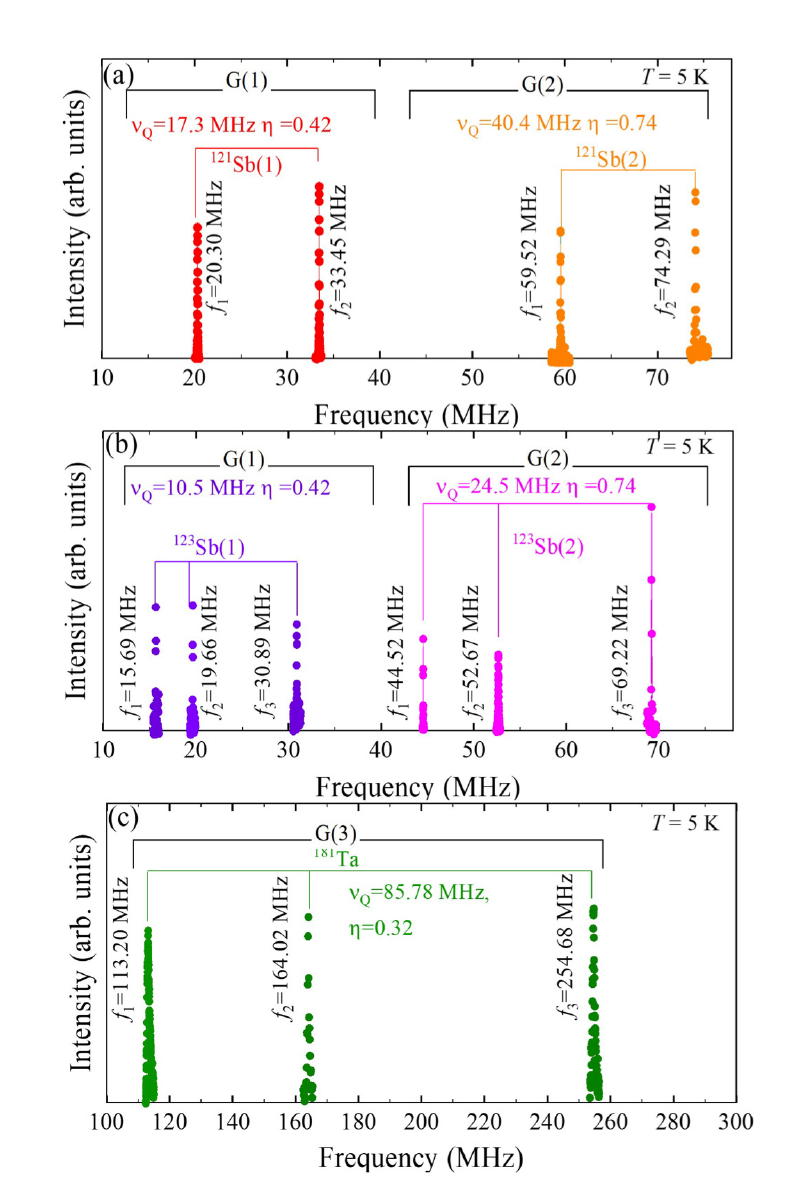}
\caption{\label{fig:nqr}
(a)$^{121}$Sb, (b)$^{123}$Sb and (c)$^{181}$Ta-NQR spectra of TaSb$_2$ at 5K. There are three groups of NQR spectra, G(1), G(2) and G(3), which are assigned to Sb(1), Sb(2) and Ta sites. The lines $f_1$, $f_2$ and $f_3$ correspond to $\pm3/2\Leftrightarrow\pm1/2, \pm5/2\Leftrightarrow\pm3/2$  and $\pm7/2\Leftrightarrow\pm5/2$ transitions, respectively.
}
\end{figure}
%--------------------------Fig.2--------------------------------------------------

Searching for the NQR signal in TaSb$_2$, we consider the splitting of nuclear energy levels due to the quadrupole interaction, which is expressed by the following Hamiltonian,
%--------------------------------------Eq.1-----------------------------%
\begin{equation}
\label{H_q}
H_Q = \frac{h\nu_Q}{6}\left[(3I{_z}^{2}-I^{2})+\frac{1}{2}\eta(I_{+}^{2}+I_{-}^{2})\right],
\end{equation}
%--------------------------------------Eq.1-----------------------------%
where $h$ is Planck’s constant, $\nu_Q$ is the nuclear quadrupole coupling constant, $\eta$ = $|V_{xx}-V_{yy}|$/$|V_{zz}|$ is the asymmetric parameter, $V_{xx}$, $V_{yy}$ and $V_{zz}$ are the EFG at nuclear sites along the $x$-, $y$-, and $z$-axis, respectively. Here, we note that the $z$-axis is the maximum principal axis of the EFG tensor, $V_{zz}$, and nuclear spins are quantized along $V_{zz}$ at zero field. The quadrupole-split nuclear energy levels $E_m$ and the resulting transition frequencies can easily be calculated numerically by diagonalizing Eq. (\ref{H_q}). For $\eta$ = 0, the energy levels can be expressed as
%--------------------------------------Eq.2-----------------------------%
\begin{equation}
\label{E_m}
E_m = \frac{h\nu_Q}{6}\left[3m^2-I(I+1)\right], \nu_Q = \frac{3eQ|V_{zz}|}{h2I(2I-1)}.
\end{equation}
%--------------------------------------Eq.2-----------------------------%
Then the NQR frequency can be written as $\nu_{NQR}$ = $\nu_Q$(2$|m|$+1)/2, because the NQR occurs for the transition between two levels $|m|$ and $|m+1|$. Therefore, for $\eta$ = 0, two and three NQR lines with equal spacings are expected for $I$ = 5/2 and $I$ = 7/2, respectively. For $\eta$$\neq$0, those NQR lines will be shifted with un-equal spacing \cite{DAS}. Hence, a total of thirteen NQR lines should be observed for $^{121}$Sb(1), $^{123}$Sb(1), $^{121}$Sb(2), $^{123}$Sb(2) and $^{181}$Ta sites in TaSb$_2$. In Fig.~\ref{fig:nqr} (a), (b) and (c), we show the observed NQR spectra at 5 K. The expected thirteen-NQR lines have clearly been observed in the frequency range of 10$\sim$260 MHz. The peak frequencies for each line have been obtained by fitting each line to a Gaussian function.

\indent The following procedures have been taken for the site assignment, namely how we can associate each line to the particular sites and nuclear isotopes. First, we consider the nuclear quadrupole moment (Q) for each nucleus to assign which site the spectrum originated from. The Q of $^{181}$Ta ($^{181}$Q = 3.170$\times$10$^{-28}$m$^2$) is approximately five times larger than that of $^{121/123}$Sb ($^{121}$Q = 0.543$\times$10$^{-28}$ and $^{123}$Q = 0.692$\times$10$^{-28}$ m$^2$). Moreover, the $^{121}$/$^{123}$Sb NQR lines at the same site are expected to be observed in a close frequency range since the ratio of $^{121}$$\nu_Q$ to $^{123}$$\nu_Q$ at the same site is 1.47 from Eq. (\ref{E_m}). Here, for the sake of clarifying, the observed NQR lines are divided into three groups according to frequency ranges of observed NQR lines and assigned as G(1) (15$\sim$40 MHz) for $^{121/123}$Sb, G(2) (40$\sim$75 MHz) for $^{121/123}$Sb, G(3) (110$\sim$255 MHz) for Ta.

\indent Next, we estimate the $\nu_Q$ from the combinations of all spectra within each group. The estimated $\nu_Q$ for $^{121}$Sb and $^{123}$Sb at each site should correspond to each other when the site-assignment is correct. Here, we also consider the asymmetry parameter $\eta$ because the peak frequencies are unequally spaced for any combinations of spectra. In order to estimate the values of $\nu_Q$ and $\eta$ from the observed peak frequencies at 5 K, a least-squares analysis was performed using theoretical quadrupole interaction obtained from the exact diagonalization. The values of $\nu_Q$ and $\eta$ have been estimated from all combinations of NQR lines within each group. We found the combinations of NQR lines in which the $\nu_Q$ and $\eta$ for $^{121}$Sb correspond to that for $^{123}$Sb, as shown in Table~\ref{tab:efg}. Therefore we conclude that G(1), G(2) and G(3) are associated with respectively Sb(1), Sb(2) and Ta sites. 

\indent The quadrupolar frequency $\nu_Q$ and the
asymmetry parameter $\eta$ can be obtained from the EFG tensors at nuclear
sites defined as the second partial derivative of the electrostatic potential
$v(r)$ at the position of the nucleus:

\begin{equation}
\label{eq:efg_dft}
V_{ij}(r=0) = \left(\frac{\partial^2{}}{\partial_i\partial_j}v(0) - \frac13\delta_{ij}\nabla^2v(0)\right),
\end{equation}
where $\delta_{ij}$ and $\nabla^2$ are the Kronecker delta and the Laplace
operator, respectively. In the DFT calculation, we choose Ta ($4f$/$5s5p6s5d6p5f$) and Sb ($4s4p4d$/$5s5d5p4f$) 
semi-core/valence states as a basis set. The inclusion of higher lying states into the basis
is crucial for an accurate estimation of the EFG tensor \cite{Yasuoka}. The
calculated $\nu_Q$ and $\eta$ are provided in Table~\ref{tab:efg} next to the
experimental values. 
Experimentally obtained values of $\nu_Q$ and $\eta$ are in good  agreement with the calculated values within 16\%. This results assure that our line assignment to the site and isotope is correct. As a side remark, we note that $\eta$ is a relative measure, and its distribution law differs from that of individual EFG components. Assuming that the latter are normally distributed, the distribution of $\eta$ can be approximated as normal, even though with the $\sim$1.5 times larger standard deviation. This difference has to be taken into account for a proper comparison.

\begin{table}[tb]
\caption{\label{tab:efg} Experimental results and calculated EFG tensor $V_{zz}$, quadrupole coupling $\nu_Q$ and asymmetry parameter $\eta$ for each site in TaSb$_2$. Sb(1) and Sb(2) are inequivalent Sb sites and the labeled one is consistent with the one named for the crystal structure.}
\begin{ruledtabular}
\begin{tabular}{cccc|ccc}
\multicolumn{3}{c}{Experiment}                                      &                      & \multicolumn{3}{c}{Calculation}                                            \\ \cline{1-3} \cline{5-7} 
\multicolumn{1}{l}{} & \multicolumn{1}{l}{} & \multicolumn{1}{l}{}  & \multicolumn{1}{l}{} & \multicolumn{1}{l}{}        & \multicolumn{1}{l}{} & \multicolumn{1}{l}{}  \\
                     & $\nu_Q$ (MHz)        & $\eta$                &                      & $V_{zz}$ (10$^{21}$V/m$^2$) & $\nu_Q$ (MHz)        & $\eta$                \\ \cline{1-3} \cline{5-7} %\hline
$^{121}$Sb(1)        & 17.3                 & \multirow{2}{*}{0.42} &                      & \multirow{2}{*}{$-7.371$}   & 14.5                 & \multirow{2}{*}{0.45} \\
$^{123}$Sb(1)        & 10.5                 &                       &                      &                             & 8.8                  &                       \\
$^{121}$Sb(2)        & 40.4                 & \multirow{2}{*}{0.74} &                      & \multirow{2}{*}{$-18.391$}  & 36.2                 & \multirow{2}{*}{0.88} \\
$^{123}$Sb(2)        & 24.5                 &                       &                      &                             & 22.0                 &                       \\
$^{181}$Ta           & 85.78                & 0.32                  &                      & $-15.147$                   & 82.9                 & 0.22                  \\
\end{tabular}
\end{ruledtabular}
\end{table}

\subsection{Weak Field NMR and determination of the principal axis}

%--------------------------Fig.3--------------------------------------------------
\begin{figure}[t]
\centering
\includegraphics[width=1\linewidth]{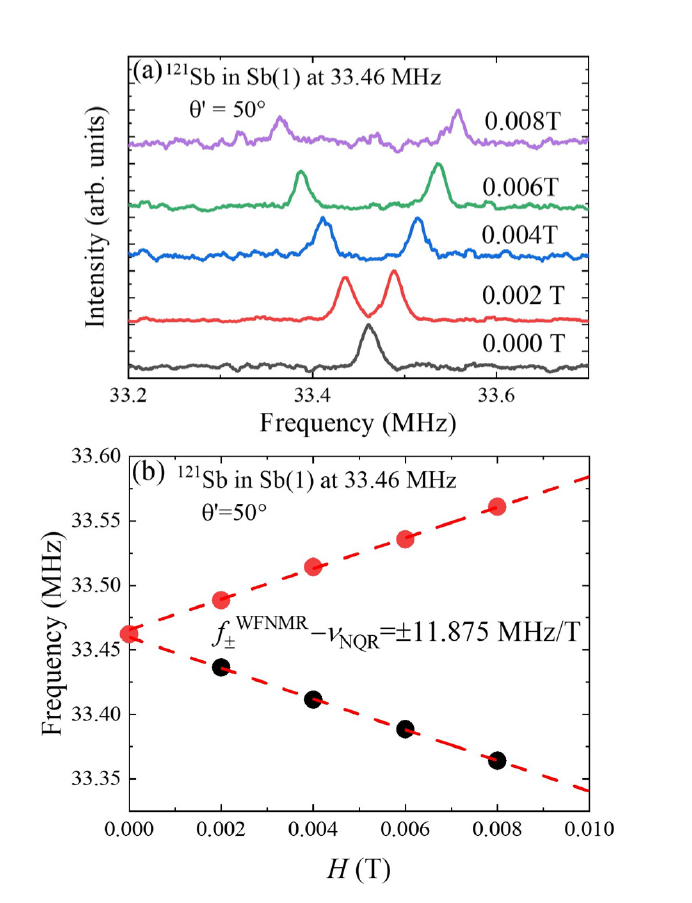}
% Here is how to import EPS art
\caption{\label{fig:wfnmr}
Typical example of the field dependence of WFNMR spectra (a) and peak frequencies (b) for $^{121}$Sb at 33.46 MHz NQR line. The weak fields were applied along $\theta'$ = 50$^{\circ}$. The slope is estimated as 11.875 MHz/T (see main text). Note that the bare gyromagnetic ratio of $^{121}$Sb is 10.189 MHz/T. 
}
\end{figure}
%--------------------------Fig.3--------------------------------------------------

To experimentally determine the principal axes of the EFG tensor, a weak magnetic field ($H$) was applied to the single crystal. Here, we can treat the Zeeman term $H_z = -\gamma_NH(I_x\sin{\theta}\cos{\phi}+I_z\cos{\theta}$) as the perturbation of $H_Q$, where $\theta$ and $\phi$ are the angle between a magnetic field and the principal axis of $V_{zz}$ in the $x-z$ plane and $x-y$ plane, respectively, and $\gamma_N$  is the bare nuclear gyromagnetic ratio. Then 
one NQR line splits into two lines ($f_{-m}^{\rm WFNMR}$ and $f_{+m}^{\rm WFNMR}$). In the case of transitions between $\pm$1/2 and $\pm$3/2, transitions between $m$ and $m$$\pm$2 also occur, so that exceptionally one NQR line splits into four lines \cite{DAS}. Therefore, generally six and eight WFNMR lines are expected for $I$ = 5/2 and 7/2, respectively (see detail in Appendix C). We define the splitting width $\Delta f = f_{-m}^{\rm WFNMR}-f_{+m}^{\rm WFNMR} = 2\gamma'H$ where $\gamma'$ is an apparent nuclei gyromagnetic ratio and the $\Delta f$ exhibits an angular dependence (see detail in Appendix C). The typical example of the WFNMR spectra at $^{121}$Sb(1) ($\nu_{NQR}$ = 33.46 MHz) is shown in Fig.~\ref{fig:wfnmr}~(a). The clear splitting of the spectrum has been observed. The peak frequencies for each magnetic field have been obtained by fitting each spectrum to two Gaussian functions. The line width of the spectra does not vary with the magnetic field and no additional peaks were observed. This fact indicates that magnetism or magnetic impurities are absent. As shown in Fig.~\ref{fig:wfnmr}~(b), the field dependence of peak frequencies is linear, and the slope is 11.875 MHz/T at $\theta'$ = 50$^{\circ}$, where $\theta'$ is an angular between the magnetic field and the axis perpendicular to the $c$-axis in the $ac$-plane. The slope greater than $\gamma$ = 10.189 is due to the effect of finite $\eta$.

\begin{figure}[t]
\centering
\includegraphics[width=1\linewidth]{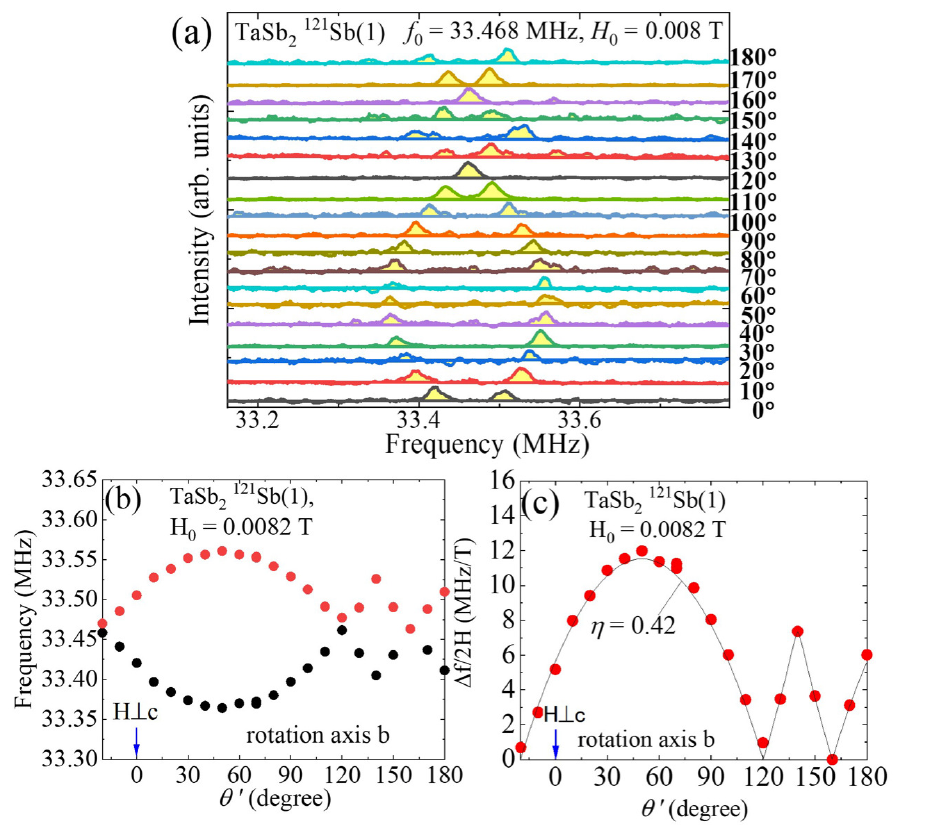}
% Here is how to import EPS art
\caption{\label{fig:angdep}
Typical example of the angular dependence of $^{121}$Sb(1) WFNMR frequency. (a) FFT spectra taken at 33.46 MHz and 0.0082 T external field. Here a single crystal was rotated along the $b$-axis. (b) Angular dependence of peak frequency of the spectra. (c) Angular dependence of $\Delta f/2H$ extracted from exact diagonalization. Solid lines are data fits to theoretical $H_Q+H_z$ from exact diagonalization where $\nu_Q$ = 17.5 MHz and $\eta$ = 0.42 were obtained.
}
\end{figure}
%--------------------------Fig.4--------------------------------------------------
\indent The principal axis of $V_{zz}$ and the associated isotope can be assigned by examining the angular dependence of $\Delta f/2H (\Delta f/2H(\theta'))$. We have measured the angular dependence of WFNMR at the highest NQR frequency lines for G(1) and G(2) and the $f_2$ line for G(3). For example, the angular dependence of WFNMR spectra and peak frequencies for $^{121}$Sb(1) are shown in Fig.~\ref{fig:angdep}~(a) and (b), respectively. Also, the angular dependence of $\Delta f/2H$ is shown in Fig.~\ref{fig:angdep}~(c). We found that the principal axis of $V_{zz}$ for Sb(1) is pointing $\theta' = 50^{\circ}$, $\phi'$ = 0 and $\eta$ = 0.42, where $\phi'$ is the angle between the principal axis of $V_{zz}$ and the $ac$-plane. Note that $\phi'$=0 means that $V_{zz}$ being within the $ac$-plane.

%by a least-squares fit of $\Delta f/2H(\theta')$ to $H_Q+H_z$ from the exact diagonalization. 
%The $\eta$ is in good agreement with the value in Table II. 
Similar measurements were made at $^{121}$Sb(2) and $^{181}$Ta, then we have found that the principal axis of $V_{zz}$ pointing $\theta'$ = 107$^{\circ}$ and 125$^{\circ}$, $\phi'$ = 0 and 0, $\eta$ = 0.74 and 0.32, respectively. The obtained $\eta$ values are consistent with the experimental values obtained from the satellite structure shown in Table \ref{tab:efg}. From the DFT calculation, we have obtained the direction of the principal axis of $V_{zz}$ as $\theta'$ = 69$^{\circ}$, 122$^{\circ}$, and 135$^{\circ}$ for Sb(1), Sb(2) and Ta sites, respectively. Therefore, the experimentally obtained direction of principal axis of $V_{zz}$ is in good agreement with the theoretical one.
%-------------------------------------SubSec. B-----------------------------%

%-------------------------------------SubSec. C-----------------------------%
\subsection{Temperature dependence of $\nu_Q$ and $\eta$}
%--------------------------Fig.5--------------------------------------------------
\begin{figure}[t]
\centering
\includegraphics[width=1\linewidth]{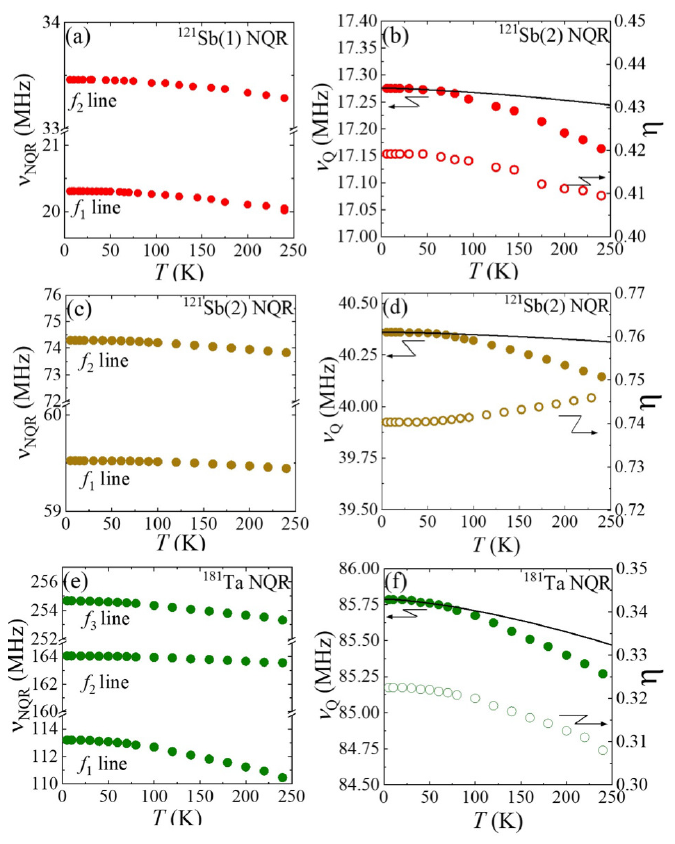}
% Here is how to import EPS art
\caption{\label{fig:tdep}
(a), (c) and (e) temperature dependences of NQR  frequency for $^{121}$Sb(1), $^{121}$Sb(2) and $^{181}$Ta, respectively. (b), (d) and (f) temperature dependence of $\nu_Q$ and $\eta$ for $^{121}$Sb(1), $^{121}$Sb(2) and $^{181}$Ta, respectively. The black solid lines represent the fits of the $\nu_Q(T)$ using $\nu_Q(0)(1-AT^{3/2})$ below 50 K. 
}
\end{figure}
%--------------------------Fig.6--------------------------------------------------
\begin{figure}[t]
\centering
\includegraphics[width=1\linewidth]{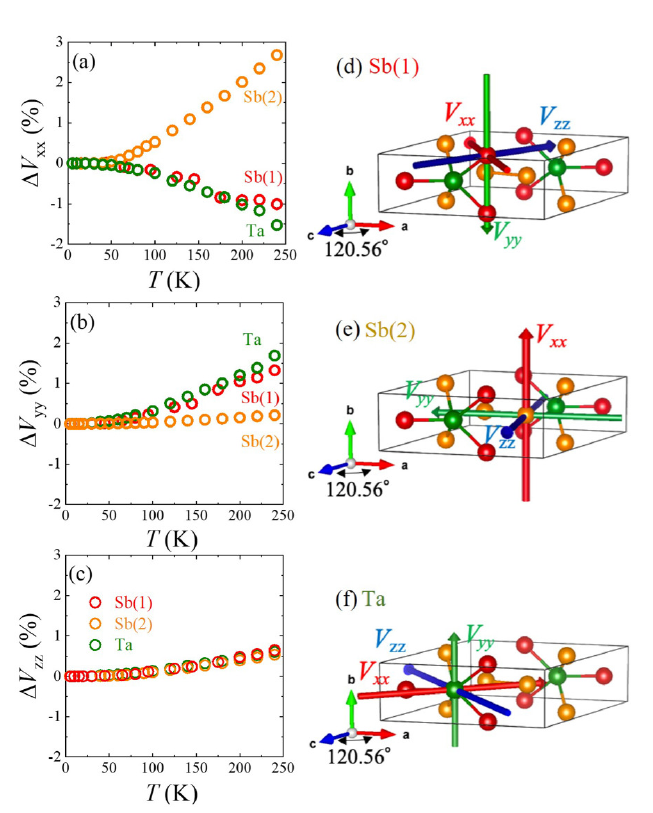}
% Here is how to import EPS art
\caption{\label{fig:fracttdep}Fractional temperature dependence of (a)$\Delta V_{xx}(T) = (V_{xx}(0)-V_{xx}(T))/V_{xx}(0)$, (b)$\Delta V_{yy}(T) = (V_{yy}(0)-V_{yy}(T))/V_{yy}(0)$ and (c) $\Delta V_{zz}(T) = (V_{zz}(0)-V_{zz}(T))/V_{zz}(0)=\Delta \nu_Q(T)$. The directions of EFG principal axis $V_{xx}$ (red arrow), $V_{yy}$ (green arrow) and $V_{zz}$(blue arrow) are illustrated for (d) Sb(1) site, (e) Sb(2) (f) Ta site.}

\end{figure}
%--------------------------Fig.6--------------------------------------------------
%--------------------------Fig.5--------------------------------------------------
In order to extract the temperature dependence of the quadrupolar parameters, $\nu_Q$ and $\eta$, the temperature dependence of the NQR frequencies for all lines have been measured up to 250 K. The results are shown in Fig.~\ref{fig:tdep}~(a), (c), and (e). The $\nu_Q (T)$ at each site follows an empirical $\nu_Q$(0)$(1-AT^{3/2})$ law below 50 K. This is a typical temperature dependence reflecting the temperature dependence of thermal expansion \cite{Chris}. From the least-square fit of $\nu_Q(T)$ to $\nu_Q(0)(1-AT^{3/2})$, we obtained the $\nu_Q(0)$ at each site, as shown in Fig. \ref{fig:tdep} (b), (d) and (f). To discuss site dependence of EFG, the fractional decrease of $V_{ii}(T)$ $(i=x, y, z)$, $\Delta V_{ii}(T) = (V_{ii}(0)- V_{ii}(T))/V_{ii}(0)$ are shown in Fig.~\ref{fig:fracttdep}~(a),(b) and (c). The deviation in $\Delta$$V_{zz}(T)$ among sites is less than 0.06\% over the whole temperature range, and there is hardly seen any site dependence on $\Delta$$V_{zz}(T)$. In contrast, $\eta$(T) show different behavior depending on the site: at the Sb(1) and Ta sites, the $\eta$(T) decreases with increasing temperature, whereas at the Sb(2) site, $\eta(T)$ increases (see Fig.~\ref{fig:tdep}~(d)).

The site-differentiation might be due to the anisotropy of the thermal expansion of lattice parameters. Here we extract $V_{xx}$ and $V_{yy}$ (see detail in Appendix D) to discuss $\eta$(T) in more detail at each site. %where the direction of $V_{zz}$ has been determined. 
%Since the EFG tensor is defined as $|V_{zz}|>|V_{yy}|>|V_{xx}|$ and $V_{zz}$+$V_{yy}$+$V_{xx}$ = 0, so $V_{xx}$ and $V_{yy}$ can be written as $V_{xx}$ = $V_{zz}$($\eta$-1)/2 and $V_{yy}$ = -$V_{zz}$($\eta$+1)/2. 
Fig.~\ref{fig:fracttdep}~(a) and (b) show the $\Delta$$V_{xx}$($T$) and $\Delta$$V_{yy}$(T) for each site, respectively. First, as a characteristic behavior, both $\Delta$$V_{xx}$ and $\Delta$$V_{yy}$ depend on temperature at the Sb(1), whereas $\Delta$$V_{yy}$ almost no temperature dependence was observed at the Sb(2). The temperature independent of $\Delta V_{yy}$ is the main source of the unusual increase of $\eta(T)$. Based on the DFT calculation, only $\Delta V_{yy}$ for Sb(2) is pointing to $a$-axis, as shown in Fig. \ref{fig:fracttdep}~(e). This result indicates that the lattice thermal expansion of the $a$-axis might be smaller than the others. Next, the $\Delta V_{yy}$ for Sb(1) and Ta and $\Delta V_{xx}$ for Sb(2), pointing to the $b$-axis, are increasing, that indicates thermal expansion induces in the $b$-axis direction.
%for Sb(1), $V_{xx}$ is located in the $ac$-plane and $V_{yy}$ is pointing along the $b$-axis, 
%for Sb(2), $V_{xx}$ is pointing to the $b$-axis and $V_{yy}$ is pointing to the $a$-axis.  
The temperature dependence of EFG tensors pointing in the $ac$-plane demonstrates that $\Delta V_{zz}$ shows an increase with increasing temperature at all sites, whereas $\Delta V_{xx}$ shows a decrease at Sb(1) and Ta. If one considers thermal expansion as the sole cause of these temperature dependencies, a contradiction arises because $\Delta V_{zz}$ indicates an expansion and $V_{xx}$ indicates a shrinkage. For a deeper insight into the mechanism of thermal expansion, low-temperature diffraction experiments and dilatometry studies are highly recommended.

%Thus, the result that only the $V_{yy}$ at Sb(2) pointing along the $a$-axis is almost independent of temperature indicates that the lattice thermal expansion of the $a$-axis might be smaller than the others. While the origin of anisotropy of thermal expansion is presently unknown, we note that one of the distinct features of the TaSb$_2$ structure is Sb(2) dimers. These dimers are aligned parallel to the $a$-axis, and their length ($\sim$ 2.77 \AA) is considerably shorter than the interatomic distances of bulk Sb ($\sim$ 2.9 \AA), possibly suggesting at strong covalent bonding. Fingerprints of the respective bonds are seen in the DOS. Indeed, if we align the local $z$-axis along the Sb(2)-Sb(2) dimer and project the DOS onto local orbitals, the bonding states at the lower edge of the valence band (about 5 eV below the Fermi energy) are dominated by $5p_z$ contributions (inset of Fig. \ref{fig:dos} (a)). 

%For a deeper insight into the mechanism of thermal expansion, low-temperature diffraction experiments and dilatometry studies are highly desirable.

\subsection{Temperature dependence of $1/T_1T$}

Before discussing the temperature dependence of $1/T_1$ based on the band structure, it is necessary to clarify that the relaxation process is governed by the magnetic excitations, rather than the quadrupolar process. One way of doing this is the careful analysis of the recovery of the nuclear magnetization from the saturated state to the thermally equilibrium state. From the recovery of nuclear magnetization, it was confirmed that the relaxation process is governed by magnetic excitation (See detail in Appendix \ref{ApB}). Furthermore, the experimentally obtained $T_1$ ratio between $^{121}$Sb and $^{123}$Sb, $^{121}$$T_1$/$^{123}$$T_1$, is a good measure of whether the relaxation process is governed by magnetic or quadrupolar, because the former case the ratio should be ($^{123}$$\gamma_N$/$^{121}$$\gamma_N$)$^2$ = 0.29 while the latter case it should be ($^{123}Q/^{121}Q)^2$ = 1.62. We have obtained the experimental ratio of 0.30 and 0.29 at Sb(1) and Sb(2), respectively. These values agree well with the  ($^{123}$$\gamma_N$/$^{121}$$\gamma_N$)$^2$. From the above two evidences, we are strongly confident that the relaxation process in TaSb$_2$ is to be discussed from magnetic excitations.

\indent Putting the experimentally obtained temperature dependence of $Q_n$ and $K_n$ due to those of $\eta(T)$ (see Fig.~\ref{fig:tdep} ~(b), (d) and (f)) to Eqs. (\ref{recovery1}) and (\ref{recovery2}) in Appendix \ref{ApB}, we have obtained the temperature dependence of $1/T_1T$ for the $f_2$ ($\pm3/2\Leftrightarrow\pm5/2$) line of $^{121}$Sb at Sb(1) and Sb(2) sites, and $f_2$ ($\pm3/2\Leftrightarrow\pm5/2$) line of $^{181}$Ta at Ta site. The obtained results are shown in Fig.~\ref{fig:t1t} where the $1/T_1T$ are divided by the square of the $\gamma_N$ to eliminate nuclear specific parameter. Although the general feature has been seen in many semi-metallic materials, one can immediately observe that there exists a characteristic temperature, $T^* \approx$ 30 K, where the relaxation process has a crossover from a high temperature activation type behavior which is presumably associated with the excitations to a higher unoccupied state to a low temperature modified Korringa excitations.

\indent Among topological materials, especially in the Dirac/Weyl fermion system with linear band dispersions, a power low behavior $1/T_1T\sim T^n$ has been observed at high temperatures  \cite{Yasuoka, Wang2020}. The power low behavior originated from a Fermi-contact and/or an orbital contribution of hyperfine interaction. When the Fermi-contact contribution is dominant, $T^4$ dependence is predicted due to the linear dispersion with the point node. While when the orbital contribution is dominant, an unusual temperature dependence of hyperfine interaction leads to a $T^2$ dependence in $1/T_1T$ \cite{Dora2016, Dora2019}. However, neither case can explain the activation-type behavior. This result is consistent with the absence of Weyl points at zero magnetic field in TaSb$_2$ from band structure calculations.

\indent Quite generally, $1/T_1T$ can be expressed by using the wave vector ($q$) and frequency ($\omega$), dependent magnetic susceptibility, $\chi(q,\omega)$, characterizing the magnetic excitations in a system as \cite{Narath1968,Moriya1963},
\begin{equation}
\label{T_1T}
\frac{1}{T_1T} = \frac{2\gamma_N^2k_B}{g^2\mu_B^2}\sum_q A_q^2 \frac{\chi_{\bot}^{''}(q,\omega)}{\omega_N},
\end{equation}
where $\chi_{\bot}^{''}(q,\omega)$ is the transverse component of imaginary part of $\chi(q,\omega)$,  $A_q$ is the $q$-dependent hyperfine coupling constant and $\omega_N$ is the frequency for the NQR. Since at the moment we do not have a plausible microscopic theory to calculate the $\chi(q,\omega)$ in multiband systems like TaSb$_2$, we have adopted the theoretical $1/T_1T$ for non-interacting itinerant electrons based on the band structure calculation with random phase approximation (RPA). For such a system $1/T_1T$ is expressed using the density of state near the Fermi level as \cite{Slichter},
%--------------------------------------Eq.9-----------------------------%
\begin{equation}
\label{T1_DOS}
\frac{1}{T_1T} \propto \frac{A_{hf}^2}{T}\int f(E-\mu_c)\left[1-f(E-\mu_c)\right]\{D(E)\}^2 dE,
\end{equation}
%--------------------------------------Eq.9-----------------------------%

\noindent where $f(E)$ is a Fermi distribution function, $D(E)$ is the energy dependent density of state and $\mu_c(T)$ is the temperature dependent chemical potential. Were there no change of $A_{hf}$ with temperature, the calculation of $1/T_1T$ is straightforward from calculated $D(E)$ (see Fig.~\ref{fig:dos}~(a)). At all sites, the calculated $1/T_1T$ shows a constant value without increase. Namely, neither the low temperature up-turn nor the high temperature activation type behaviors are reproduced by the calculated $1/T_1T$ based on the DOS.
The absence of sharp DOS features in the immediate vicinity of the Fermi level
excludes the possibility of sizable shifts of the $\mu_c$ (and
hence the DOS at $\mu_c$) at low temperatures. Thus, the nature of the
experimentally observed upturn in $1/T_1T$
remains elusive in the band structure. One possible explanation is that the DFT fails to describe the
electronic structure of TaSb$_2$ accurately. While deficiencies of DFT can not
be excluded especially for highly covalent systems, we have two arguments that
render such a scenario unlikely. First, our DFT calculations accurately
describe the EFG tensors. Second, our results are in agreement with the linear
coefficient $\gamma=1.2$\,mJ$\cdot$mol$^{-1}\cdot$K$^{-2}$ estimated from the
Sommerfeld expansion of the specific heat at low temperatures (see Appendix E).  Indeed,
$\gamma$ can be estimated directly from the DOS at the Fermi energy
$D_{\varepsilon_{\text{F}}}$ (in J$^{-1}$ per formula unit)

\begin{equation}
\label{eq:sommerfeld}
\gamma_{\text{DFT}} \left[\frac{\text{mJ}}{\text{mol}\cdot{}\text{K}^2}\right] = \frac{k_{\text{B}}^2 \pi^2 N_{\text{A}}V_{\text{uc}}}{3} {D_{\varepsilon_{\text{F}}}} \times 10^3,
\end{equation}

\noindent where $k_{\text{B}}$ is the Boltzmann constant (in J/K), $N_{\text{A}}$ is the
Avogadro number (in mol$^{-1}$), $Z=2$ is the number of formula units per cell,
and $V_{\text{uc}}$ is the unit cell volume (in m$^3$).  In the full
relativistic calculations, the resulting $\gamma_{\text{DFT}}$ values amount to
1.00$\pm$0.05\,mJ$\cdot$mol$^{-1}\cdot$K$^{-2}$. We also simulated the very small possible deviation of the sample stoichiometry from the ideal 1:2 ratio (see Appendix \ref{ApA}) applying the virtual crystal approximation and found no significant changes in the DOS near the Fermi level.
%, depending on the structural input. 
The agreement between the calculated and the experimental values also not only supports our DFT analysis, but also indicates that band renormalization in TaSb$_2$ is small. 
The discrepancy between the experimental $1/T_1T(T)$ and those from the band structure calculation demonstrate that TaSb$_2$ is not a simple semi-metal, and the magnetic excitation has to be treated beyond the simple Fermi liquid scenario.
%\indent Here, we note that TaSb$_2$ may not be a simple semimetal, but the possibility that the activation type behavior of $1/T_1T$ is a unique excitation due to the Dirac/Weyl fermions is excluded. In Dirac/Weyl semimetals, a power low behavior of $1/T_1T(T)$ were observed \cite{Yasuoka, Wang2020} and the power low behavior originated by a Fermi-contact and/or an orbital contribution of hyperfine interaction. When the Fermi-contact contribution is dominant, $T^4$ dependence is predicted due to the linear dispersion including the line node. While when the orbital contribution is dominant, an unusual temperature dependence of hyperfine interaction leads to a $T^2$ dependence in $1/T_1T$ \cite{Dora2016, Dora2019}. 

%and results are shown by dashed curves in Fig.~\ref{fig:t1t}, where the local density of states for Sb(1), Sb(2) and Ta sites were implemented to Eq. (\ref{T1_DOS}). As clearly observed in the figures, all calculated results indicate that there exists no essential temperature dependence of $1/T_1T$. While the calculated results is consistent with the linear temperature dependence of the electron specific heat at low temperatures (see Sec. V). The discrepancy between the experimental $1/T_1T(T)$ and those from the band structure calculation as well as macroscopic measurements demonstrate that TaSb$_2$ is not a simple semi-metal, and the magnetic excitation has to be treated beyond the RPA.  \\

\begin{figure}[t]
\centering
\includegraphics[width=1\linewidth]{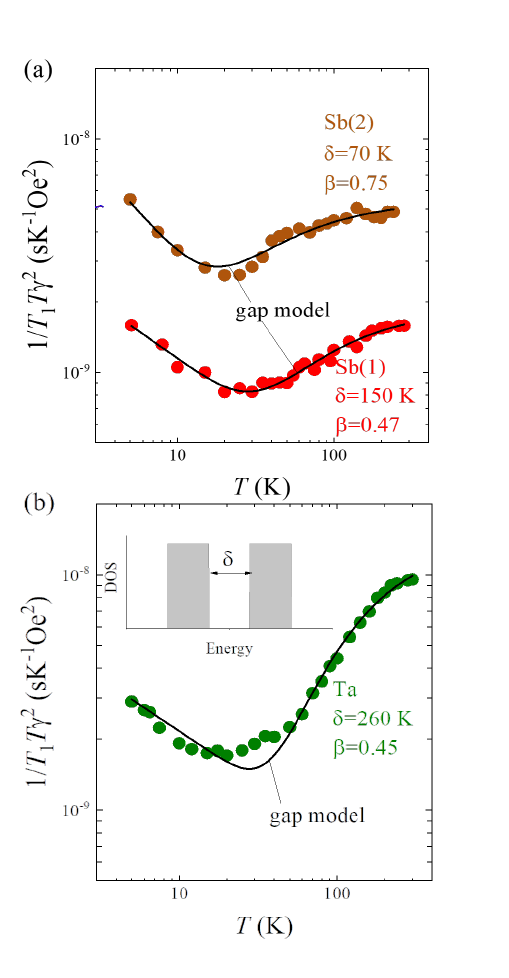}
% Here is how to import EPS art
\caption{\label{fig:t1t}
Temperature dependence of $1/T_1T\gamma_N^2$ for (a) $^{121}$Sb at Sb(1), Sb(2) and (b) $^{181}$Ta. The solid lines are the data fits to Eq.~\ref{eq:gapmodel} with parameter $\delta$ = 150 K, 70 K and 260 K for Sb(1), Sb(2) and Ta, respectively. inset: the model of density of states with gap $\delta$. 
%The dashed curves are calculated $1/T_1T\gamma_N^2$ followed by  Eq. (\ref{T1_DOS}) with partial DOS for each site (see Fig.~\ref{fig:dos}~(a)). Here, the absolute values of calculated $1/T_1T\gamma_N^2$ are normalized by $(1/T_1T)_{\infty}$ estimated from data fits for each site.
}
\end{figure}

\indent  Following a number of previous NMR studies like SmB$_6$ \cite{Takigawa1981}, PuB$_{4}$ \cite{Dioguardi2019} and FeGa$_3$ \cite{Gippius2014}, we assume an existence of rather arbitral in-gap state  which is invisible in calculated band structure to analyze the $1/T_1T$ . Here, $D(E)$ exhibits a symmetric band gap $\delta$ centered at $E_F$ shown in inset of Fig. \ref{fig:t1t} (b).
The origin of the narrow band in-gap state is not clear but may be associated with the Anderson localization or impurity states.
Assuming a simple in-gap state with narrow bands which is invisible in the calculated band structure, the activation-type temperature dependence of $1/T_1T (T)$ is discussed by a phenomenological way. The data are then fitted to the following empirical equation, including the upturn at low temperatures,

\begin{equation}
\label{eq:gapmodel}
%\begin{split}
\frac{1}{T_1T}=\alpha T^{-\beta}+\left(\frac{1}{T_1T}\right)_0\exp\left(-\frac{\delta}{2k_BT}\right)
%\end{split}
\end{equation}

The first term describes a fluctuation of the localized electron moments within the narrow band at low temperatures, and the second term describes thermal excitation of the quasiparticles.
The solid lines in Fig.~\ref{fig:t1t}~(a) and (b) indicate such fit of the $1/T_1T(T)$ to Eq.~\ref{eq:gapmodel} with $\delta$ = 13 meV(150K), 6 meV(70K) and 23 meV(260 K) for Sb(1), Sb(2) and Ta sites are obtained, respectively. 
%Such a small energy scale in the activation process has been observed in previous NMR study on SmB$_6$ ($\delta$ = 4.3 meV) \cite{Takigawa1981}, FeGa$_3$ ($\delta$ = 1.1 meV) \cite{Gippius2014} and PuB$_4$ ($\delta$ = 1.8 meV) \cite{Dioguardi2019} and no real explanation has been drawn so far. 
The obtained different activation energies at each site might indicate the presence of site dependent gap. As indicated by the site dependence of the gap size, the increase of $1/T_1T$ with temperature at the Ta site is larger than at the Sb site. Such differences in site-specific behavior have also been observed in Weyl semimetal TaAs \cite{Kubo}. The origin of the site dependence is unknown, but a theoretical study is warranted.

%Following the common phenomenological treatment of $1/T_1T(T)$ at high temperature region, an activated-type temperature dependence, $1/T_1T=(1/T_1T)_{\infty} \exp(-\delta/2k_BT)$, has been applied to each site, where $\delta$ is the activation energy of magnetic excitations and 
%$(1/T_1T)_{\infty}$ is the constant value at which thermal excitation exceeds the thermal activation energy and enters the Korringa regime. 

\indent 
The power law coefficient $\beta$ in the first term is associated with the strength of exchange coupling between lower occupied state where $\beta$=1 and $\beta$=0 correspond to purely localized state and delocalized state, respectively. The fit value of $\beta$$\sim$0.5, averaged for all sits in TaSb$_2$, as well as for TaAs \cite{Kubo}, may indicate that the assumed in-gap state is rather spatially extended so that the exchange narrowed theory which predicts $\beta$ = 1 may not be applicable. This explanation is on the assumption of in-gap states associated with the Anderson localized or impurity state as a phenomenological description, and the following possibilities for the origin of the upturn can also be mentioned. i) Contribution from the mobility edge near the gapped band \cite{Schlottmann2014}, ii) Fluctuations in the Kondo state (impurity induced Kondo hole) \cite{Caldwell} iii) Spin diffusion to the magnetic impurity in topological insulator \cite{Shishiuchi2002}. Nevertheless, the idea that the nuclear spin diffusion to dilute impurity moments is discussed in YbB$_{12}$ \cite{Shishiuchi2002} seems to be one of the plausible interpretations. Likewise, the similar mechanism from bulk to surface metallic state in topological materials is worth investigating from theoretical and experimental viewpoints.

\section{Concluding Remarks}
%We have reported characteristic static and dynamical properties of observed thirteen NQR lines associated with $^{121}$Sb, $^{123}$Sb and $^{181}$Ta in TaSb$_2$. 
The presented study is a combination of experimental nuclear quadrupole resonance (NQR) and theoretical calculations using crystal structure obtained from our x-ray single crystal diffraction. We have reported characteristic static and dynamical properties of observed thirteen NQR lines associated with $^{121}$Sb, $^{123}$Sb and $^{181}$Ta in TaSb$_2$. All $^{121/123}$Sb and $^{181}$Ta NQR lines are successfully assigned to all inequivalent sites and isotopes by combined analysis of the DFT calculation and the WFNMR measurement. The following characteristic features are to be noted. 

\indent(I) From the data fitting of the temperature dependence of NQR lines the zero-temperature quadrupole coupling constant $\nu_Q$ and asymmetry parameter $\eta$ are extracted, and those values are in agreement with the DFT calculations within $\pm$16$\%$. The temperature dependence of $\nu_Q (T)$ and $\eta(T)$  has been measured up to 250 K, and unusual $\eta(T)$ was observed for the Sb(2) site. The angular dependence of WFNMR frequencies for all sites enabled to determine the direction of the EFG principal axes, $V_{xx}$ and $V_{yy}$ and $V_{zz}$ with respect to the monoclinic crystal axes. The temperature dependencies of the $V_{xx}(T)$ and $V_{yy}(T)$ and $V_{zz}(T)$ have shown a unique characteristic where we found the $V_{yy}(T)$ of Sb(2) site has almost temperature independent. Since the $V_{yy}$ of Sb(2) is pointing exactly to the $a$-axis, the origin of the unexpected $\eta(T)$ may be attributed to the small lattice expansion of the $a$-axis.

\indent(II) The nuclear relaxation process measured in the NQR at all the sites is proved experimentally by the magnetic excitations, not the quadrupolar. The temperature dependence of relaxation rate, $1/T_1T(T)$, has quite characteristic behavior similar to the many semimetal materials. That is the high temperature activation type process crosses over to the low temperature excess relaxation process, making an upturn in $1/T_1T(T)$ around 30-70 K. However, neither high nor low temperature behaviors could be explained by the simple Fermi liquid using the DFT density of states at Fermi level, even the temperature dependence of the chemical potential is included. Followed by the traditional procedure, the activation type behavior of $1/T_1T(T)$ was analyzed by assuming simple in gap state could reproduce high temperature $1/T_1T(T)$. This energy scale of the activation is similar to other topological materials. We did not observe any power-law behavior in $1/T_1T(T)$ characteristic to Weyl or Dirac fermion excitations. \\
\indent Our DFT-supported NQR/NMR study with the local probes Ta and Sb provide with a valuable contribution to understanding of the temperature dependent local charge distribution, band structure and magnetic excitations in the topological semimetal TaSb$_2$.

\begin{acknowledgments}
OJ and HR thank U. Nitzsche for technical assistance. We thank V. Hasse for technical support in the characterization and synthesis of the samples.
\end{acknowledgments}

%\bibliography{reference}
%\bibliographystyle{plainnat}
%\bibliographystyle{apsrev4-2}

\appendix
\section{Sample characterization (EDX and XRD)}
\label{ApA}
The chemical composition of the as-grown TaSb$_2$ crystals have been determined by energy dispersive X-ray spectroscopy (EDXS) in a scanning electron microscope (Jeol JSM 7800F). Measurements with an acceleration voltage of 22 kV result in an analytical totals of more than 97 mass $\%$ by using the ZAF matrix correction model for the calculation of the mass concentration from the intensities of the X-ray lines Sb L$\alpha$ (3.6 keV) and Ta L$\alpha$ (8.1 keV). The respective element ratio does not deviate significantly from formal concentration of Ta: 33.3 at. $\%$, Sb: 66.6 at.$\%$. X-ray intensities are measured with the silicon drift detector (Bruker Esprit ver.2.3 EDX system) attached to the SEM.
\begin{figure}[H]
\centering
\setcounter{figure}{0}
\renewcommand{\thefigure}{A\arabic{figure}}
\includegraphics[width=1\linewidth]{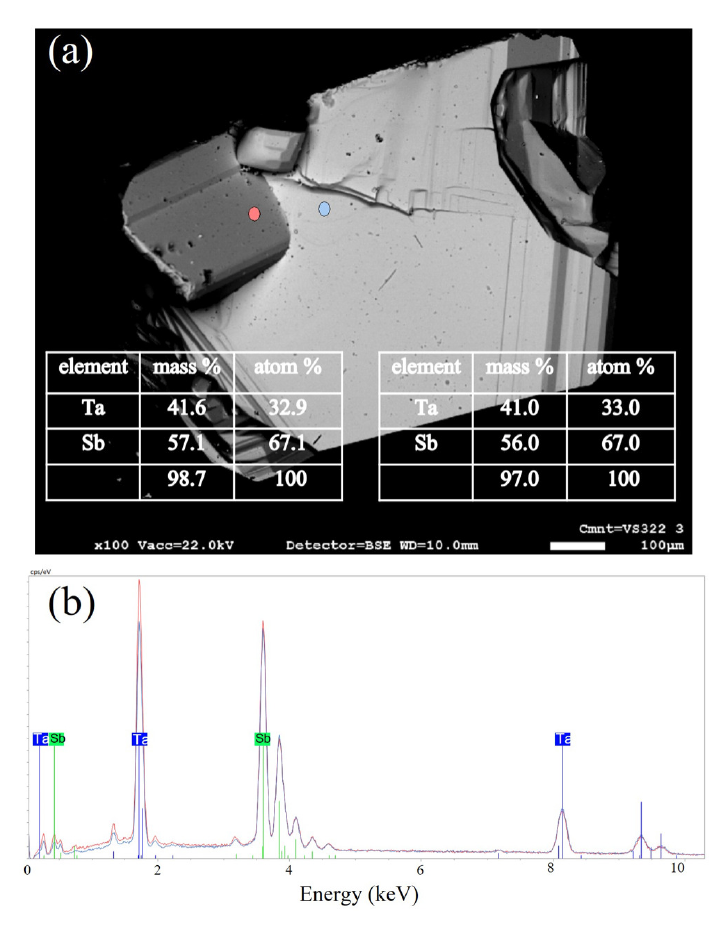}
\caption{\label{fig:D1}
(a) BSE (back scattered electron) image of as grown crystals. The chemical composition (see Tables) are determined by semi-qualitative EDX method. (b) Energy-dispersive x-ray spectra (color refers to electron beam position in (a)).
}
\end{figure}
The crystal structure investigation of TaSb$_2$ was carried out by using X-ray single crystal diffraction. Suitable specimen for the investigation was mechanically extracted from larger crystal grown by chemical transport. Intensity data collection was performed by using Rigaku AFC7 diffractometer equipped by Saturn 724+ CCD detector (MoK$\alpha$ radiation). Structural model with one Ta and two Sb crystallographic positions (all in 4i Wyckoff sites) was refined in the space group $C2/m$ ($a$ = 10.233(2) \AA, $b$ = 3.6487(5) \AA, $c$ = 8.3036(19) \AA, $\beta$ = 120.40(1) $^{\circ}$, $R1$ = 0.028, $wR2$ = 0.067). In separate cycle of the refinement procedure, the occupancy factors of all atomic positions were allowed to vary freely. This showed no deviation from the ideal values, confirming the 1 : 2 stoichiometry of the examined sample. The final runs were implemented with the full (fixed) occupancies for all atomic positions (see Tables \ref{table:D2}, \ref{table:D3}). The least squares fit of the experimental X-ray powder pattern to that calculated from the obtained structural model showed good agreement (Fig. \ref{fig:ddx}), indicating high purity of the synthesized material.

\begin{figure}[H]
\centering
\setcounter{figure}{1}
\renewcommand{\thefigure}{A\arabic{figure}}
\includegraphics[width=1\linewidth]{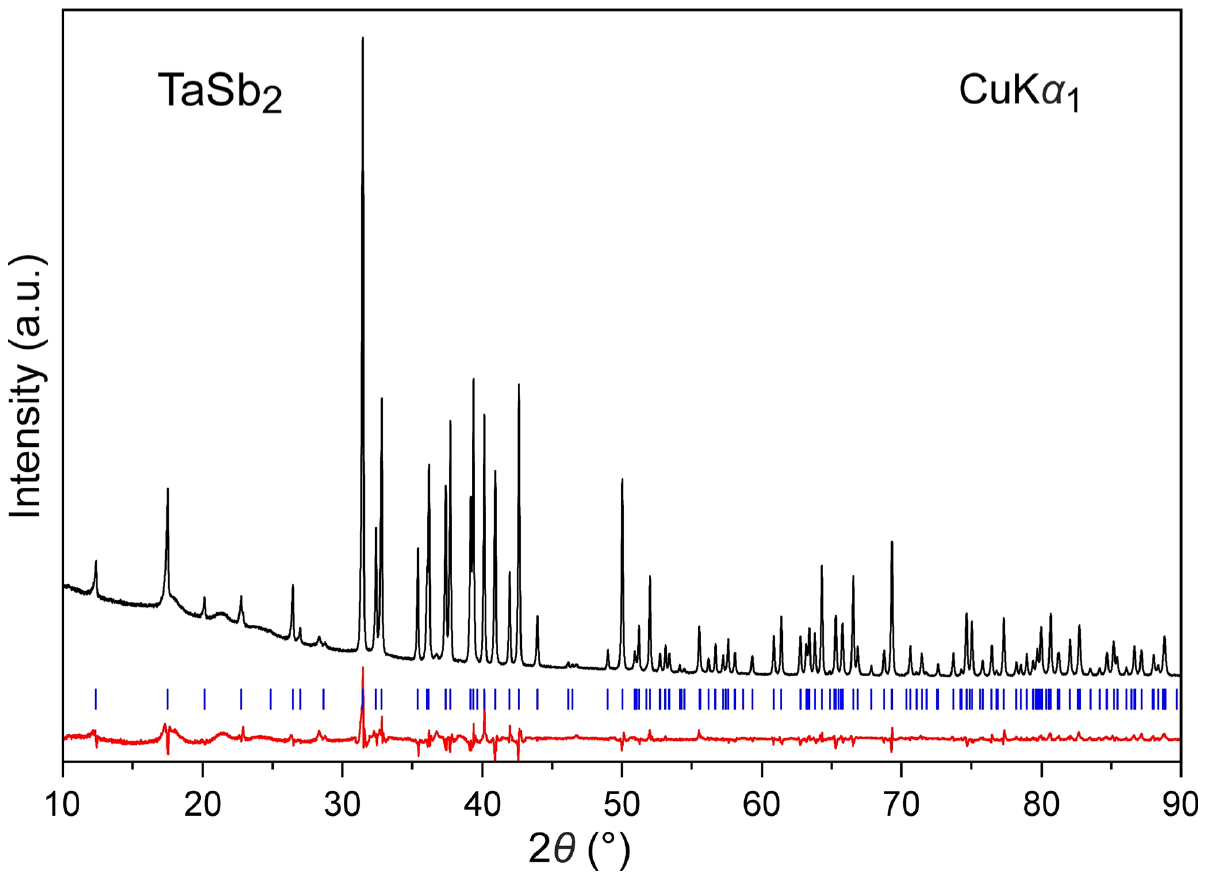}
\caption{\label{fig:ddx}
X-ray powder diffraction recorded in Guinier geometry (black line). The difference between measured and calculated intensities is represented as a red curve. Blue vertical bars indicate reflection positions of the refined model. 
}
\end{figure}

\begin{table}[H]
\centering
\setcounter{table}{0}
\renewcommand{\thetable}{A\arabic{table}}
\begin{tabular}{ll}
\hline
Composition                                 & TaSb$_2$                        \\
Space group                                 & $C2/m$                          \\
Pearson symbol                              & $mS12$                          \\
Formula units per unit cell, Z              & 4                               \\
Lattice parameters                          &                                 \\
$a$ / \AA                    & 10.233(2)                       \\
$b$ / \AA                    & 3.6487(5)                       \\
$c$ / \AA                    & 8.3036(19)                      \\
$\beta$ / $^{\circ}$                         & 120.40(1)                       \\
$V$ / \AA$^3$                & 267.41(9)                       \\
Cal. density / gcm$^{-1}$                   & 10.54                           \\
Crystal form                                & irregular shaped                \\
Crystal size / $\mu$m                       & 0.030$\times$0.035$\times$0.060 \\
Diffraction system                          & RIGAKU AFC7                     \\
Detector                                    & Saturn 742+CCD                  \\
Radiation, $\lambda$/\AA      & Mo$K\alpha$, 0.71073            \\
Scan; step / degree; N(images)              & $\phi$, 0.8, 900                \\
Maximal 2$\theta$ /degree                   & 78.0                            \\
Range in $h, k, l$                          & -18 $\leq$ $h$ $\leq$ 17          \\
                                            & -6 $\leq$ $k$ $\leq$ 4            \\
                                            & -14 $\leq$ $l$ $\leq$ 10          \\
Absorption correction                       & multi-scan                      \\
$T$(max)/$T$(min)                           & 0.453                           \\
Absorption coeff. / mm$^{-1}$               & 60.5                            \\
$N(hkl)$ measured                           & 2509                            \\
$N(hkl)$ unique                             & 847                             \\
R$_{int}$                                   & 0.041                           \\
$N(hkl)$ observed                           & 812                             \\
Observation criteria                        & $F(hkl)$ $\geq$ 4$\sigma$(F)    \\
Refined parameters                          & 20                              \\
$R1$                                        & 0.028                           \\
$wR2$                                       & 0.067                           \\
Residual peaks / e\AA$^{-3}$ & -5.00/4.68                      \\ \hline
                                            &                                
\end{tabular}
\caption{\label{table:D1}Crystallographic data for TaSb$_2$.}
\end{table}

\begin{table}[H]
\centering
\setcounter{table}{1}
\renewcommand{\thetable}{A\arabic{table}}
\begin{tabular}{c|c|c|c|c|l}
Atom  & site & $x/a$      & $y/b$ & $z/c$      & $U_{eq}$    \\ \hline
Ta    & 4i   & 0.15151(3) & 0     & 0.18914(3) & 0.00492(11) \\ \hline
Sb(1) & 4i   & 0.40541(5) & 0     & 0.11318(6) & 0.00578(11) \\ \hline
Sb(2) & 4i   & 0.14706(5) & 0     & 0.53391(6) & 0.00602(11)
\end{tabular}
\caption{\label{table:D2}Atomic coordinates and equivalent displacement parameters (in \AA$^2$) in the crystal structure of TaSb$_2$.
}
\end{table}

\begin{table}[H]
\centering
\setcounter{table}{2}
\renewcommand{\thetable}{A\arabic{table}}
\begin{tabular}{c|c|c|c|c}
Atom  & $U_{11}$      & $U_{22}$      & $U_{33}$      & $U_{13}$      \\ \hline
Ta    & 0.00531(14) & 0.00481(15) & 0.00437(15) & 0.00225(10) \\ \hline
Sb(1) & 0.00657(19) & 0.00509(19) & 0.0062(2)   & 0.00361(15) \\ \hline
Sb(2) & 0.00621(19) & 0.00608(19) & 0.00562(19) & 0.00289(15)
\end{tabular}
\caption{\label{table:D3}Anisotropic displacement parameters (in \AA$^2$) in the crystal structure of TaSb$_2$. $U_{12}=U_{23}=0$.
}
\end{table}

\section{Recovery of nuclear magnetization}
\label{ApB}

For the magnetic relaxation, the time dependence of the recovery of nuclear magnetization measured by NQR intensity should obey Eqs. (\ref{recovery1}) and (\ref{recovery2}) for $I$ = 5/2 ($^{121}$Sb) and $I$ = 7/2 ($^{123}$Sb and $^{181}$Ta), respectively \cite{Chepin1991}.

%--------------------------Fig.7--------------------------------------------------
\begin{figure}[H]
\centering
\setcounter{figure}{0}
\renewcommand{\thefigure}{B\arabic{figure}}
\includegraphics[width=1\linewidth]{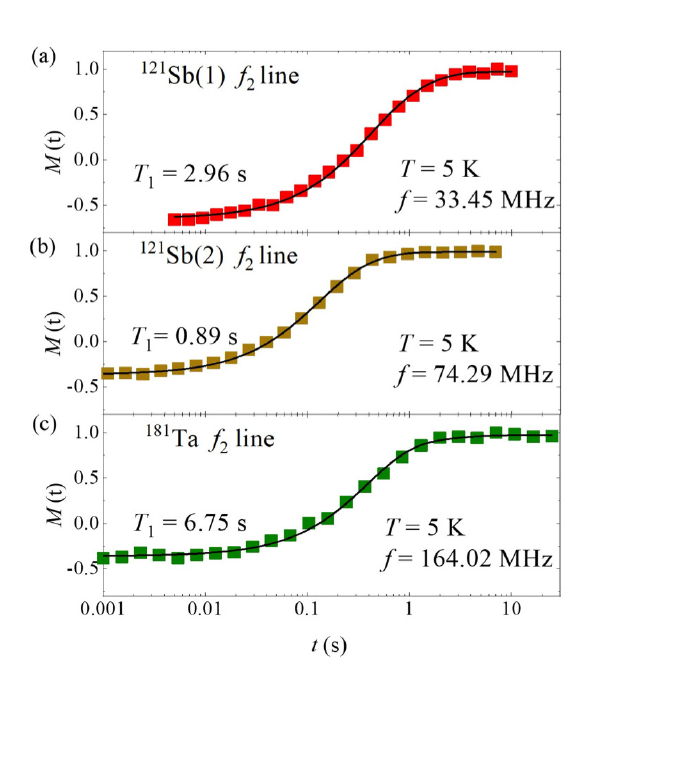}
% Here is how to import EPS art
\caption{\label{fig:mrecover}
Recovery of nuclear magnetization $M (t)$ measured by the spin-echo amplitude after an inversion pulse for $^{121}$Sb ((a) and (b)) and $^{181}$Ta ((c)) NQR lines at 5K. The black solid lines are data fits to Eqs. (\ref{recovery1}) and (\ref{recovery2}) for  $^{121}$Sb and $^{181}$Ta, respectively.  
}
\end{figure}
%--------------------------Fig.7-------------------------------------------------
\begin{table}[H]
\setcounter{table}{0}
\renewcommand{\thetable}{B\arabic{table}}
\caption{\label{tab:qnkn}The calculated prefactors $Q_n$ and exponent $K_n$ in Eqs. (\ref{recovery1}) and (\ref{recovery2}) for $\eta = 0.42, \eta = 0.74$ and $\eta = 0.32$ for $^{121}$Sb(1), $^{121}$Sb(2) and $^{181}$Ta \cite{Chepin1991}.}
\begin{ruledtabular}
\begin{tabular}{cccccc}
\multirow{4}{*}{$^{121}$Sb(1)} & \multicolumn{5}{c}{$K_1$ = 3.0215,  $K_2$ = 8.9861}                                                                  \\ \cline{2-6} 
                               & \multicolumn{1}{l}{} & \multicolumn{1}{l}{} & \multicolumn{1}{l}{} & \multicolumn{1}{l}{} & \multicolumn{1}{l}{} \\
                               & $f_1$ line           &                      & $Q_1$ = 0.1432         & $Q_2$ = 0.8568         &                      \\
                               & $f_2$ line           &                      & $Q_1$ = 0.3783         & $Q_2$ = 0.6217         &                      \\ \cline{2-6} 
\multicolumn{1}{l}{}           & \multicolumn{1}{l}{} & \multicolumn{1}{l}{} & \multicolumn{1}{l}{} & \multicolumn{1}{l}{} & \multicolumn{1}{l}{} \\

\multirow{4}{*}{$^{121}$Sb(2)} & \multicolumn{5}{c}{$K_1$ = 3.0444,  $K_2$ = 8.1479}                                                                  \\ \cline{2-6} 
                               & \multicolumn{1}{l}{} & \multicolumn{1}{l}{} & \multicolumn{1}{l}{} & \multicolumn{1}{l}{} & \multicolumn{1}{l}{} \\
                               & $f_1$ line           &                      & $Q_1$ = 0.1972         & $Q_2$ = 0.8028         &                      \\
                               & $f_2$ line           &                      & $Q_1$ = 0.3085         & $Q_2$ = 0.6915         &                      \\ \cline{2-6} 
\multicolumn{1}{l}{}           & \multicolumn{1}{l}{} & \multicolumn{1}{l}{} & \multicolumn{1}{l}{} & \multicolumn{1}{l}{} & \multicolumn{1}{l}{} \\
\multirow{5}{*}{$^{181}$Ta}    & \multicolumn{5}{c}{$K_1$ = 3.0645,  $K_2$ = 9.2297,  $K_3$ = 18.2449}                                                  \\ \cline{2-6} 
                               & \multicolumn{1}{l}{} & \multicolumn{1}{l}{} & \multicolumn{1}{l}{} & \multicolumn{1}{l}{} & \multicolumn{1}{l}{} \\
                               & $f_1$ line           &                      & $Q_1$ = 0.0441         & $Q_2$ = 0.4015         & $Q_3$ = 0.5544         \\
                               & $f_2$ line           &                      & $Q_1$ = 0.0862         & $Q_2$ = 0.0584         & $Q_3$ = 0.8554         \\
                               & $f_2$ line           &                      & $Q_1$ = 0.2032         & $Q_2$ = 0.5526         & $Q_3$ = 0.2442         \\
 \cline{2-6} 
\end{tabular}
\end{ruledtabular}
\end{table}

%--------------------------------------Eq.6,7-----------------------------%
\begin{multline}
\label{recovery1}
M_{I = 5/2}=M_0\left[1-c_0\left(Q_1\exp{\frac{-K_1t}{T_1}}\right.\right.\\
\left.\left.+Q_2\exp{\frac{-K_2t}{T_1}}\right)\right],
\end{multline}
\begin{multline}
\label{recovery2}
M_{I = 7/2}=M_0\left[1-c_0\left(Q_1\exp{\frac{-K_1t}{T_1}}\right.\right.\\
\left.\left.+Q_2\exp{\frac{-K_2t}{T_1}}+Q_3\exp{\frac{-K_3t}{T_1}}\right)\right],
\end{multline}
%--------------------------------------Eq.6,7-----------------------------%
where $c_0$ is the degree of inversion, and $Q_n$ ($n$ = 1,2,3) and $K_n$ ($n$ = 1,2,3) are the parameters depending on the nuclear spin $I$ and $\eta$. All parameters are calculated in Ref. \cite{Chepin1991} for a given $I$ and $\eta$. The typical examples are shown in Fig.~\ref{fig:mrecover}, where the experimental recovery curves for the $f_2$ lines in $^{121}$Sb(1), $^{121}$Sb(2) and $^{181}$Ta NQR lines are shown with the best fit curve to Eqs. (\ref{recovery1}) and (\ref{recovery2}) using $K_n$ and $Q_n$ shown in Table B1. The obtained perfect fit provides us with the relaxation process is governed by the magnetic fluctuations.

\section{Angular dependence of $\Delta f$}
\begin{figure}[H]
\centering
\setcounter{figure}{0}
\renewcommand{\thefigure}{C\arabic{figure}}
\includegraphics[width=1\linewidth]{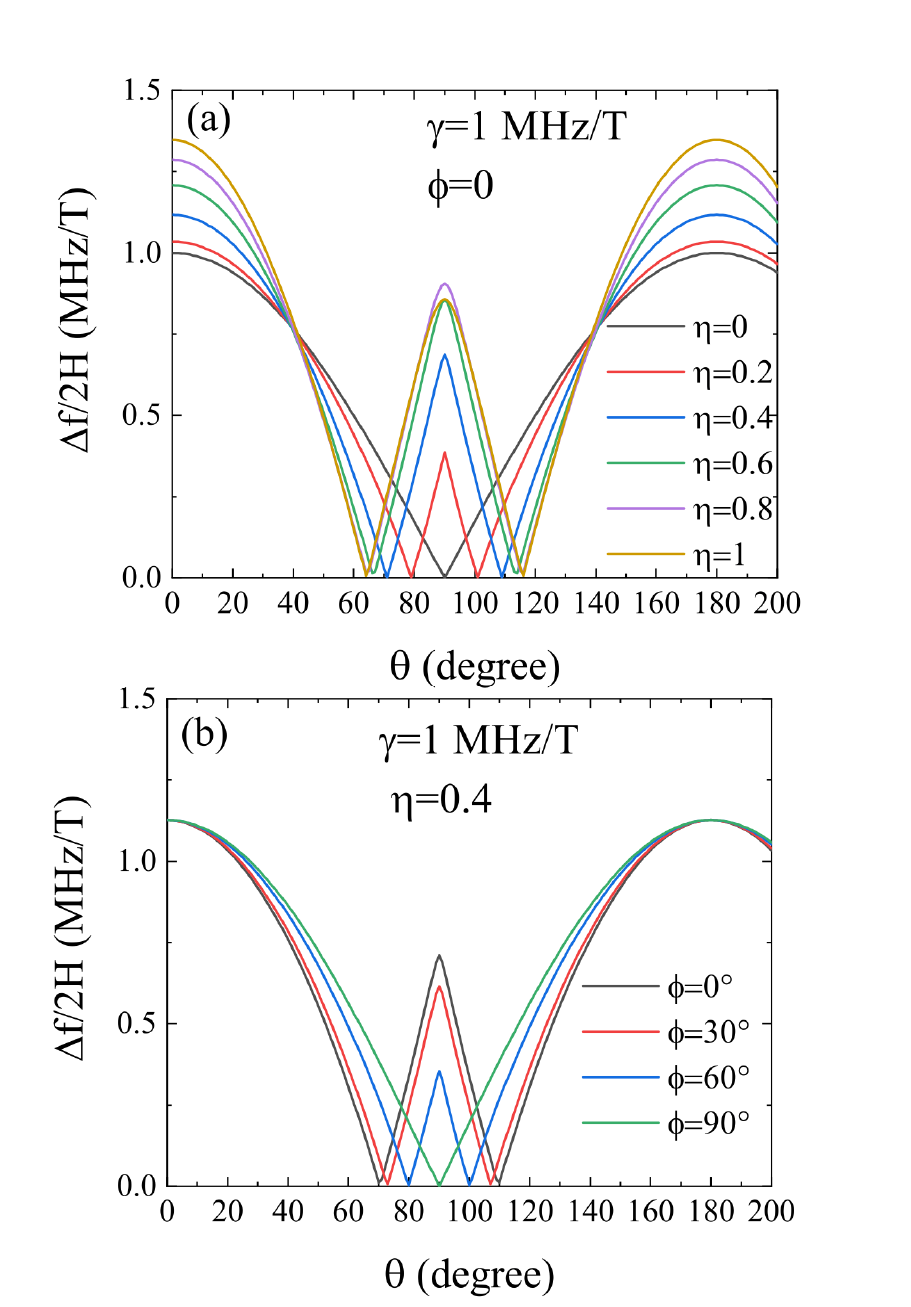}
\caption{\label{fig:ed}
Angular dependence of $\Delta f/2H$ calculated using the exact diagonalization of Eqs. (\ref{m_Hami1}), (\ref{m_Hami2}) and (\ref{m_Hami3}) (a) for $\eta$ = 0$\sim$ 1 (b) $\phi$=0$\sim$90 degree.}
\end{figure}
When we apply $H$ to the sample, we treat the Zeeman term $H_z = -\gamma_NH(I_x\sin{\theta}\cos{\phi}+I_z\cos{\theta}$) as the perturbation of $H_Q$. Then the Hamiltonian is modified as follows,

\begin{multline}
\label{m_Hami1}
\Bra{m}H_Q+H_z\Ket{m} = \left[-m \hbar \gamma_N H\cos{\theta}+\frac{e^{2}Q}{6I(2I-1)} \right.\\
\left[\{3m^2-I(I+1)\}\{V_{zz}-\frac{1}{2}(V_{xx}-V_{yy})\}\right]\Braket{m|m},
\end{multline}
\begin{multline}
\label{m_Hami2}
\Bra{m\pm1}H_Q+H_z\Ket{m} = \mp\frac{1}{2}\gamma_N H\\
\sqrt{I(I+1)-m(m\pm1)}\sin{\theta}\cos{\phi}\Braket{m|m},
\end{multline}
\begin{multline}
\label{m_Hami3}
\Bra{m\pm2}H_z+H_Q\Ket{m} = \left[\frac{e^{2}Q}{6I(2I-1)}
\left[\frac{3}{4}(V_{xx}-V_{yy})\right.\right.\\
\left.\left.\sqrt{I(I+1)-m(m\pm1)}\sqrt{(I+1)-(m\pm1)(m\pm2)}\right]\right]\\
\Braket{m|m}.
\end{multline}

%--------------------------------------Eq.3,4,5-----------------------------%
We obtain nuclear energy levels, $E_m$, by the Zeeman effect resulting from the exact diagonalization of matrix which consists of Eqs. (\ref{m_Hami1}), (\ref{m_Hami2}) and (\ref{m_Hami3}). When $\eta$ = 0 and $\theta$ = 0 ($H$ $\parallel$ $V_{zz}$), WFNMR occurs for the transition between two levels $m$ and $m\pm$1 ($|m|$$>$3/2), and the resonance frequency can be expressed as, $f_{\pm m}^{WFNMR} = \nu_{NQR}\pm\gamma_NH$.

Following Eqs. (\ref{m_Hami1}), (\ref{m_Hami2}) and (\ref{m_Hami3}), we know that a measurement of the angular dependence of $\Delta f/2H$ makes it possible to determine the principal axis direction of $V_{zz}$ ($\theta$ and $\phi$) and $\eta$. We have calculated the angular dependence of $\Delta f/2H$ using the exact diagonalization of Eqs. (\ref{m_Hami1}), (\ref{m_Hami2}) and (\ref{m_Hami3}). Here, we assume that $V_{zz}$ is along $\theta$ = 0$^{\circ}$, $\phi$ = 0$^{\circ}$ and $\gamma$ = 1 MHz/T. The maximum of $\Delta f/2H$ increases with $\eta$, as shown in Fig. \ref{fig:ed}~(a). And the $\Delta f/2H$ at 90$^{\circ}$ also changes with $\eta$. On the other hand, Fig. \ref{fig:ed}~(b) shows that $\phi$ has no effect on the maximum value of $\Delta f/2H$, but the 90$^{\circ}$ component decreases with $\phi$. Therefore, the direction of $V_{zz}$ and $\eta$ can be obtained by analyzing the experimental results of the angular dependence of $\Delta f/2H$ using a least-squares analysis with $\eta$ and $\phi$ as parameters.
%\begin{figure}[]
%\setcounter{figure}{0}
%\renewcommand{\thefigure}{A\arabic{figure}}
%\includegraphics[width=0.9\linewidth]{Fig_A1}
%%\includegraphics[width=1\linewidth]{Fig_Ab}% Here is how to import EPS art
%\caption{\label{fig:ed}
%Angular dependence of $\Delta f/2H$ calculated using the exact diagonalization of Eqs. (\ref{m_Hami1}), (\ref{m_Hami2}) and (\ref{m_Hami3}) (a) for $\eta$ = 0$\sim$ 1 (b) $\varphi$=0$\sim$90 degree.}
%\end{figure}

\section{Temperature dependence of EFG tensor}
To examine the EFG tensor in detail, $V_{xx}$, $V_{yy}$ and $V_{zz}$ were obtained by decomposing the temperature dependence of $\nu_Q(T)$ and $\eta(T)$. The $V_{zz}(T)$ was determined based on Eq.~\ref{E_m} in the main text, and the sign was determined by the DFT calculation (see in Table \ref{tab:efg}). Since the EFG tensor is defined as $|V_{zz}|>|V_{yy}|>|V_{xx}|$ and $V_{zz}$+$V_{yy}$+$V_{xx}$ = 0, $V_{xx}$ and $V_{yy}$ can be written as $V_{xx}$ = $V_{zz}$($\eta$-1)/2 and $V_{yy}$ = -$V_{zz}$($\eta$+1)/2. Thus we determined $V_{xx}(T)$, $V_{yy}(T)$, and $V_{zz}(T)$ for each site are shown in Fig. D1(a$\sim$ i).

\begin{figure}[]
\setcounter{figure}{0}
\renewcommand{\thefigure}{D\arabic{figure}}
\includegraphics[width=1\linewidth]{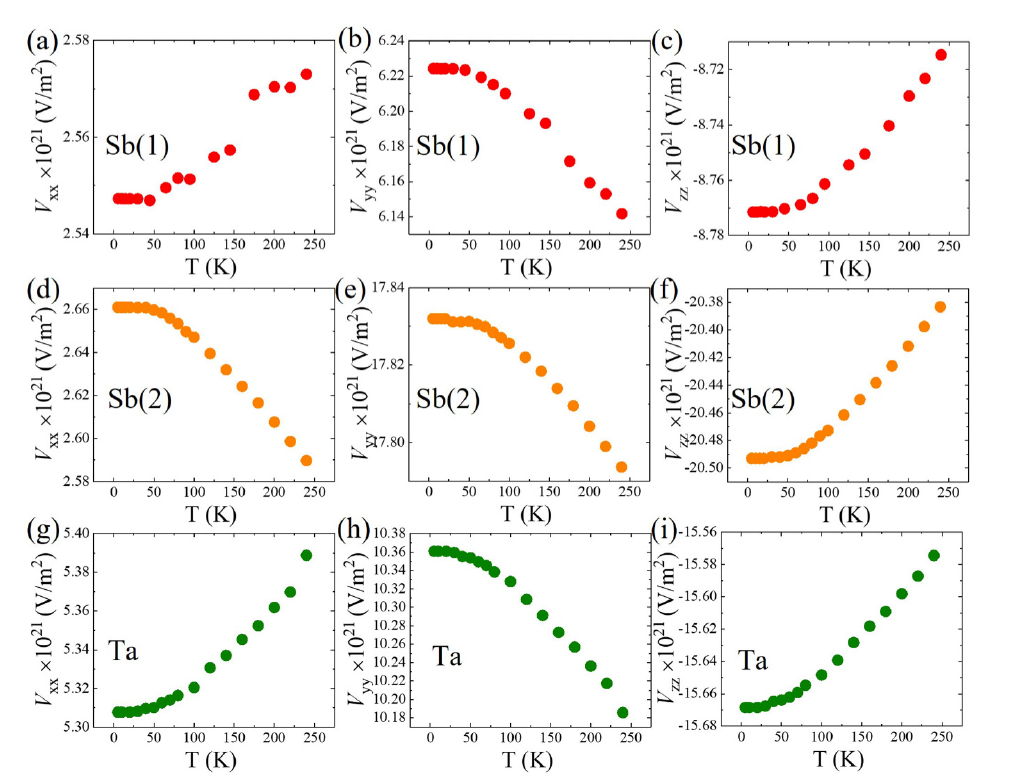}
\caption{\label{fig:defg}
Temperature dependence of EFG tensor $V_{xx}$, $V_{yy}$ and $V_{zz}$ at (a)(b)(c) Sb(1), (d)(e)(f) Sb(2) and (g)(h)(i) Ta site.}
\end{figure}

\section{Specific heat}
The specific heat of a TaSb$_2$ single crystal was measured in the temperature range between 2 K and 300 K using Quantum Design PPMS system (Fig. \ref{fig:dp}). At high temperatures, the Dulong Petit law was proven. At low temperatures, the data can be described with a linear (electron) and a cubic (phonon) dependence. It is clear that i) there is no increase in $C/T$ which means the increase in $1/T_1T$ at low temperatures may not be electron origin and ii) the determined value for the linear part is 1.2 $\rm{mJ/molK^2}$ which is within the range of what is estimated from the band structure calculations (Fig.~\ref{fig:dos} b in the main text).
%We have also measured the temperature dependence of specific heat using single
%crystal sample to evaluate the density of states from view of the macroscopic
%measurement, as shown in Fig. \ref{fig:dp} (a). Resulting from the fits $C=\gamma T+\beta T^3$ to temperature dependence of specific heat, the $\gamma$ value is obtained as 1.2 $\rm{mJ/molK^2}$.

\begin{figure}[H]
\centering
\setcounter{figure}{0}
\renewcommand{\thefigure}{E\arabic{figure}}
\includegraphics[width=1\linewidth]{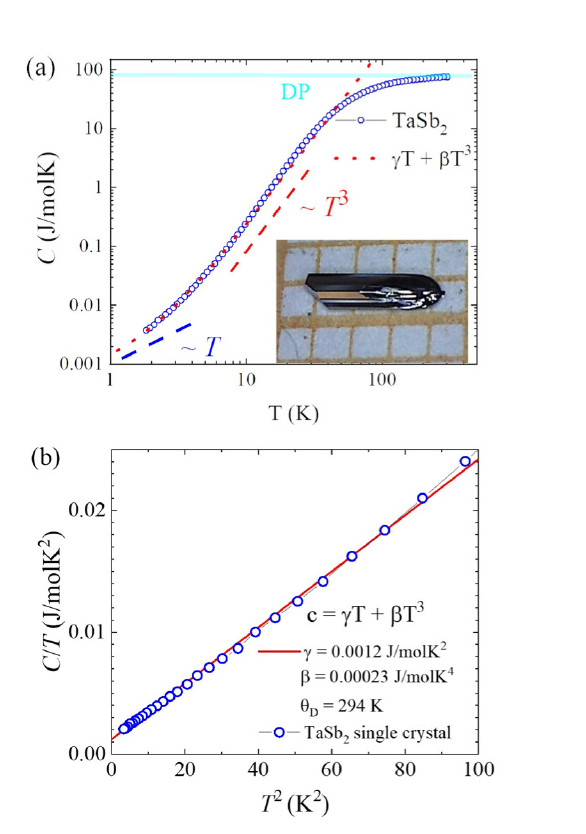}
\caption{\label{fig:dp}
(a) Specific heat as a function of temperature. DP denotes the Dulong Petit value. The curved dotted line indicates a description of the data below 10 K with electron and phonon contribution. (b) Plot of the specific heat $C/T$ versus $T^2$. The straight line indicates a description with electron and phonon contribution. The intercept at $T=0$ indicates an electronic contribution to the specific heat.
}
\end{figure}

\end{document}